\documentclass[11pt,letterpaper,usenames,dvipsnames]{article}

\usepackage[T1]{fontenc}
\usepackage[latin9]{inputenc}
\usepackage{longtable}
\usepackage[authoryear]{natbib}
\usepackage[caption=false]{subfig}
\usepackage{comment}
\usepackage{ae,aecompl,geometry,float,framed,rotfloat,amsthm,soul,amsmath,
amssymb,graphicx,todonotes,setspace,enumitem,
booktabs,rotating,microtype,soul,units,bbm,mathtools, comment}
\usepackage{xcolor}
\usepackage[hyphens]{url}
\usepackage{dsfont}
\usepackage{hyperref}
\hypersetup{colorlinks=true,citecolor=Bittersweet,linkcolor=Bittersweet,urlcolor=Bittersweet}

\makeatletter

\newcommand*\diff{\mathop{}\!\mathrm{d}}

\theoremstyle{plain}

  \theoremstyle{plain}
  \newtheorem{lem}{\protect\lemmaname}
  \theoremstyle{definition}
  
  \theoremstyle{plain}
  
  \theoremstyle{plain}
  \newtheorem{prop}{\protect\propositionname}
  \theoremstyle{plain}
  \newtheorem*{cor*}{\protect\corollaryname}
\newcounter{asscount}
\usepackage{amsthm}

\newtheoremstyle{assumption}
  {0.2cm}{0cm}
  {\rmfamily}
  {0cm}
  {\bfseries}{ }
  {0cm}
  {\thmname{#1}\thmnumber{ #2}:\thmnote{ #3}}
\theoremstyle{assumption}
\newtheorem{ass}{Assumption}

\newtheoremstyle{prediction}
  {0.2cm}{0cm}
  {\rmfamily}
  {0cm}
  {\bfseries}{ }
  {0cm}
  {\thmname{#1}\thmnumber{ #2}:\thmnote{ #3}}
\theoremstyle{prediction}
\newtheorem{pred}{Prediction}

\@ifundefined{showcaptionsetup}{}{%
 \PassOptionsToPackage{caption=false}{subfig}}
\usepackage{subfig}
\makeatother
  \providecommand{\corollaryname}{Corollary}
  \providecommand{\definitionname}{Definition}
  \providecommand{\lemmaname}{Lemma}
  \providecommand{\propositionname}{Proposition}
\providecommand{\theoremname}{Theorem}

\newcommand{\T}{\textbf{T}}

\usepackage{titling}

\newcommand{\qmg}{q_\text{t}}

\newcommand{\LT}{\textbf{LT}}
\newcommand{\LP}{\textbf{LP}}
\newcommand{\A}{\textbf{A}}

\renewcommand{\citet}[1]{\cite{#1}}

\usepackage[normalem]{ulem}

\begin{document}

\newcommand{\titlepaper}{Fragmentation and optimal liquidity supply on decentralized exchanges 
}

\onehalfspacing  

\title{\textbf {\huge \titlepaper}}

\newcommand{\keywords}{FinTech, decentralized exchanges (DEX), liquidity, fragmentation, adverse selection}
\newcommand{\JELcodes}{G11, G12, G14}

\newcommand{\mazabstract}{\noindent 

\noindent 


We investigate how liquidity providers (LPs) choose between high- and low-fee trading venues, in the face of a fixed common gas cost. Analyzing Uniswap data, we find that high-fee pools attract 58\% of liquidity supply yet execute only 21\% of volume. Large LPs dominate low-fee pools, frequently adjusting out-of-range positions in response to informed order flow. In contrast, small LPs converge to high-fee pools, accepting lower execution probabilities to mitigate adverse selection and liquidity management costs. Fragmented liquidity dominates a single-fee market, as it encourages more liquidity providers to enter the market, while fostering LP competition on the low-fee pool.

\bigskip{}
}

\author{ Alfred Lehar\\ Christine A. Parlour\\  Marius Zoican\thanks{Alfred Lehar (\href{alfred.lehar@haskayne.ucalgary.ca}{alfred.lehar@haskayne.ucalgary.ca}) is affiliated with Haskayne School of Business at University of Calgary. Christine A. Parlour (\href{parlour@berkeley.edu}{parlour@berkeley.edu}) is with Haas School of Business at UC Berkeley. Marius Zoican (\href{marius.zoican@rotman.utoronto.ca}{marius.zoican@rotman.utoronto.ca}, corresponding author) is affiliated with University of Toronto Mississauga and Rotman School of Management. Corresponding address: 3359 Mississauga Road, Mississauga, Ontario L5L 1C6, Canada. We have greatly benefited from discussion on this research with Michael Brolley, Agostino Capponi (discussant), Itay Goldstein, Sang Rae Kim (discussant), Olga Klein, Katya Malinova (discussant), Ciamac Moallemi, Uday Rajan, Thomas Rivera, Andreas Park, Gideon Saar, Lorenzo Sch\"{o}enleber (discussant), Andriy Shkilko (discussant), and Shihao Yu. 
We are grateful to 
conference participants at the 2024 NYU Stern Market Microstructure Meeting, Tokenomics 2023, Gillmore Centre Annual Conference 2023, Edinburgh Economics of Technology, Financial Intermediation Research Society 2023, the Northern Finance Association 2023, the UNC Junior Faculty Finance Conference, as well as to 
seminar participants at the University of Chicago, Lehigh, the Microstructure Exchange, UCSB-ECON DeFi Seminar, University of Melbourne, Wilfrid Laurier University, University of Guelph, Rotman School of Management, Hong Kong Baptist University, and Bank of Canada.
Marius Zoican gratefully acknowledges funding support from the Rotman School of Managements' FinHub Lab and the Canadian Social Sciences and Humanities Research Council (SSHRC) through an Insight Development research grant (430-2020-00014).
}}


\maketitle

\vspace{-10mm}


\begin{abstract}
\mazabstract


\noindent \textbf{JEL Codes}: \JELcodes

\thispagestyle{empty}

\newpage{}
\thispagestyle{empty}
\end{abstract}

\vfill{}

\vfill{}

\pagebreak{}

\vspace*{20mm}
\begin{center}
\huge \titlepaper
\par\end{center}{\Large \par}

\vspace{12mm}

\bigskip{}
\begin{abstract}
\mazabstract

\bigskip{}


\noindent \textbf{JEL Codes}: \JELcodes
\bigskip{}

\newpage{}
\setcounter{page}{1}
\end{abstract}

\newpage
\setcounter{page}{1}

\section{Introduction \label{sec:introduction}}

In addition to aggregating information, asset markets allow agents to exhaust private gains from trade. While there is a well developed literature on the informativeness of prices, less is known about if and when markets effectively exhaust all gains from trade.\footnote{Gains from trade comprise an idiosyncratic private value for the underlying asset, but also idiosyncratic preferences for trade speed or ``liquidity.''Agents' idiosyncratic value for the underlying asset are plausibly determined by their portfolio positions, and therefore independent of the trading place.  By contrast, their idiosyncratic preference for liquidity determines market quality.} In this paper, we exploit the unique design of a decentralized exchange to shed light on the market for liquidity and show, theoretically and empirically, that market fragmentation can improve trading efficiency.  

Automated market makers such as Uniswap v3 provide a unique environment to investigate the market for liquidity. While there are various new institutional details that animate these exchanges, for our purposes  three are economically important.   First, in automated exchanges, liquidity demand and supply can easily be distinguished:  users either supply or demand liquidity.  Because of this, we can isolate the effect of transactions costs on each side of the market for liquidity.  Second, costs and benefits incurred by liquidity suppliers are easier to observe because prices are not set by market participants, but are automatically calculated as a function of liquidity demand and supply. Thus, liquidity suppliers are not compensated through price impact. Third, market participants are pseudo-anonymous so we can identify and document liquidity suppliers at a high frequency. These unique features allow us to investigate, theoretically and empirically, how transactions costs affect liquidity supply. 

 Beyond investigating the market for liquidity, there are three additional reasons to investigate liquidity provision in AMMs.  First, these markets are large and successful in their own right: After its May 2021 launch, Uniswap v3 features daily trading volume in excess of  US \$1 billion. Second,  for major pairs such as Ether against USD stablecoins, Uniswap boasts twice or three times better liquidity than continuous limit order exchanges such as Binance, which suggests that this design can be economically superior.\footnote{See \href{https://uniswap.org/blog/uniswap-v3-dominance}{The Dominance of Uniswap v3 Liquidity}; May 5, 2022.} Third, as traditional assets become tokenized, and markets become more automated, this new market form could be adopted.\footnote{Swarm --- a BaFin regulated entity --- already offers AMM trading for a variety of tokenized Real World Assets.}

Uniswap v3 provides two innovations over the previous v2. First, liquidity suppliers and demanders  select into  trading places (called pools) that differ on transaction fees. Each asset pair to be traded on up to four liquidity pools that only differ in the compensation for liquidity providers: in particular, liquidity fees can be equal to 1, 5, 30, or 100 basis points and the corresponding tick sizes are 1, 10, 60, or 200 basis points.  These proportional fees are paid by liquidity demanders and are the only source of remuneration to liquidity providers.  (These fees, as we discuss below, are similar to the make-take fees that are prevalent in limit order markets.) 
Second, on Uniswap v3, liquidity providers can submit ``concentrated liquidity."  Even though their liquidity is passively supplied, they can choose the price range over which it is supplied. With volatile assets, these concentrated liquidity positions can become stale and require rebalancing. 

Besides differences in fees, the liquidity pools are otherwise identical and, importantly, they share the common infrastructure of the Ethereum blockchain.  Importantly,  all participants pay a transaction cost (called a ``gas fee'') to access the markets. Our theory and empirical work investigates the effect of different proportional fees and fixed fees on liquidity supply. At launch, Uniswap Labs conjectured that trading and liquidity should consolidate in equilibrium on a single ``canonical'' pool for which the liquidity fee is just enough to compensate the marginal market maker for adverse selection and inventory costs. That is, activity in low-volatility pairs such as stablecoin-to-stablecoin trades should naturally gravitate to low fee liquidity pools, whereas speculative trading in more volatile pairs will consolidate on high fee markets.\footnote{See \emph{Flexible fees} paragraph at \url{https://uniswap.org/blog/uniswap-v3}; accessed September 14, 2022.} As we show, this reasoning is flawed. 

We present a simple model with trade between liquidity suppliers and two types of liquidity demand. Consistent with the design of v3, liquidity suppliers  chose a market and then place their liquidity into a band around the current value of the asset. The posted liquidity is subject to a bonding curve and hence generates a price impact cost for the liquidity demanders (we emphasize that this does not remunerate the liquidity suppliers).  Liquidity suppliers have heterogeneous endowments, interpretable as different capital constraints --- low-endowment liquidity providers are akin to retail traders, whereas high-endowments stand in for large institutional investors or quantitative funds. Trade occurs against these positions either because a liquidity demander arrives who has experienced a liquidity shock or because the value of the asset has changed and an arbitrageur adversely selects the passive liquidity supply.  Collectively, the decisions of the liquidity demanders determine the payoff to the liquidity suppliers. After large private or common value trades, liquidity providers rebalance their positions; to do so, liquidity providers incur a fixed cost (i.e., gas price) each time they update their position.

Traders demanding liquidity face two types of costs:  first, the fee associated with their chosen pool (low or high) and second, the price impact costs generated by the pool's bonding curve and supplied liquidity. We find that traders route small orders exclusively to the low-fee pool to obtain the all-in lowest cost. In contrast, large traders split their orders across both low- and high-fee liquidity pools. As a result, low-fee markets are actively traded and require frequent liquidity updates whereas high-fee pools have a longer liquidity update cycle since they absorb fewer trades.

We  establish conditions under which there is fragmentation or consolidation.  Specifically, even in this simple framework, there is a robust parameter range in which liquidity does not naturally concentrate on one of the exchanges. 
Both pools can attract a positive market share if liquidity providers face gas fees and the adverse selection costs are sufficiently low. Liquidity providers trade off a higher revenue per unit of time in the low-fee pool (driven by the larger trading volume) against higher adverse selection as well as the additional gas cost required for active liquidity management. As a result, liquidity provider clienteles emerge in equilibrium. Liquidity providers with large endowments gravitate towards low-fee markets, as they are best positioned to frequently update their position. In contrast, smaller market makers choose to passively provide liquidity on high-fee markets where they only trade against large orders being routed there. They optimally trade off a lower execution probability against higher fees per unit of volume, reduced adverse selection, as well as a lower liquidity management cost per unit of time. 

Not only does liquidity fragment, but it differs in both use and type across the two markets. A small number of highly active large liquidity providers, potentially institutional investors and hedge funds, primarily trade against numerous small incoming trades on pools with low fees. In contrast, high-fee pools involve less frequent trading between a substantial number of capital-constrained passive liquidity providers (e.g., retail market makers) on one side and a few sizeable incoming orders on the other. As the fixed gas fee affects liquidity providers pool choice, 
  changes in the common fixed market access fee differentially affects the liquidity supply on the two pools.  Specifically, it reduces market quality (in the sense of lower posted liquidity) on the low fee pool. 

As we distinguish between liquidity demanders who are trading to exploit gains from trade and liquidity demanders who are arbitraging common value changes, we can decompose returns to liquidity providers, and show that adverse selection is higher on the low fee pool.  Given that the low fee pool is populated with larger liquidity suppliers, this suggests that institutional traders bear price risk.

Our findings indicate that liquidity fragmentation can enhance market quality, as measured by total gains from trade. In a single fee market, a fee that is too low fails to attract liquidity providers with smaller endowments and thus more sensitive to fixed costs, leading to unrealized gains. Conversely, a very high fee results in prohibitively high trading costs and deters trade. A two-pool market with heterogeneous fees offers two instruments to independently manage costs. The higher fee determines the marginal liquidity provider $\LP$ entering the market, and therefore the realized gains from trade. The lower fee pool, by attracting $\LP$ with larger endowments, can reduce transaction costs. We demonstrate that a two-pool fee structure can always be designed to yield higher gains from trade than any single-pool arrangement.

Using the model for guidance, we analyze  more than 28 million interactions with Uniswap v3 liquidity pools -- that is, all liquidity updates and trades from the inception of v3 in May 2021 until July 2023. 
We first document  liquidity fragmentation in 32 out of 242 asset pairs in our sample, 
which account for 95\% of liquidity committed to Uniswap v3 smart contracts and 93\% of trading volume. For each of the fragmented pairs, trading consolidates on two pools with adjacent fee levels: either 1 and 5 basis points (e.g., USDC-USDT), 5 and 30 basis points (ETH-USDC), or 30 and 100 basis points (USDC-CRV).

We then document that high-fee pools are on average larger -- with aggregate end-of-day liquidity of \$46.50 million relative to \$33.78 million, the average size of low-fee pools. Nevertheless, three quarters of daily trading volume executes on low-fee pools. In line with the model predictions, low-fee pools are more active as they capture many small trades. There are five times as many trades on low- than on high-fee pools (610 versus 110). However, the average trade on the high fee pool is twice as large: \$14,490 relative to \$6,340. Unsurprisingly, liquidity cycles -- measured as the time between the submission and update of posted liquidity -- are 20\% shorter on the highly active low-fee pool. 

We find robust evidence of liquidity supply clienteles across pools. The average liquidity deposit is 107.5\% larger on the low-fee pool, after controlling for daily volume and return volatility. At the same time, high-fee pools' market share is 21 percentage points higher. The results point to an asymmetric match between liquidity supply and demand: large liquidity providers are matched with small liquidity demanders on low-fee pools, whereas small liquidity providers trade with a few large orders on the high-fee pool. 

We then turn to the common fixed cost of accessing the market, or gas fees.  The market shares of the liquidity pools depend on the magnitude of gas costs on the Ethereum blockchain. In the model, a higher gas price leads to a shift in liquidity supply from the low- to the high-fee pool as active position management becomes relatively more costly for the marginal liquidity provider. We find that a one standard deviation increase in gas prices corresponds to a 4.63 percentage points decrease in the low-fee pool market share, and a 29\% drop in liquidity inflows on days when gas costs are elevated. 

Consistent with our model, we find that liquidity providers in low-fee pools earn higher fee yields but face increased adverse selection costs. Specifically, the daily fee yield is 2.03 basis points larger on low-fee pools. On the other hand, the permanent price impact as measured by loss-versus-rebalancing \citep[LVR, as in ][]{zhang2023amm} is 6.39 basis points or 81\% greater in low-fee pools compared to high-fee ones. However, despite this difference, the deviations in prices between high- and low-fee pools and those on centralized exchanges do not differ significantly.



Our paper is related to various literatures. 
 \citet{pagano1989trading} shows that if an asset is traded on two identical exchanges with equal transaction costs, in equilibrium market participants gravitate to a single exchange due to network effects. In practice, exchanges are rarely identical: fragmentation can emerge between fast and slow exchanges \citep{pagnotta2018competing, brocim:20} or between lit and dark markets \citep{Zhu2014}. In our model, fragmentation on decentralized exchanges is driven by variation in liquidity fees as well as different economies of scale due to heterogeneity in liquidity provider capital. We find that liquidity fragmentation driven by high gas fees implies larger transaction costs on incoming orders. We note that there is no time priority on decentralized exchanges, which clear in a pro rata fashion. On markets with time priority, \citet{FoucaultMenkveld2008} and \citet{Ohara2011market} find that market segmentation in equity markets improves liquidity (by allowing queue jumping) and price discovery.

Fixed costs for order submission are uncommon in traditional markets. However, in 2012, the Canadian regulator IIROC implemented an ``integrated fee model'' that charged traders for all messages sent to Canadian marketplaces. \citet{KorajczykMurphy2018} document that this measure disproportionately affected high-frequency traders, resulting in wider bid-ask spreads but lower implementation shortfall for large traders, possibly due to a reduction in back-running activity. Our study contributes additional insights by highlighting that the introduction of a fixed cost, even when applied across exchanges, can lead to market fragmentation.

We also relate to a rich literature on market fragmentation and differential fees. Closest to our paper, \citet{Battalio2016} and \citet{Cimon2021} study the trade-off between order execution risk and compensation for liquidity provision in the context of make-take fee exchanges. However, \citet{Battalio2016} specifically addresses the issue of the broker-customer agency problem, whereas our study focuses on liquidity providers who trade on their own behalf. In traditional securities markets, make-take fees are contingent on trade execution and proportional to the size of the order.  On the other hand, gas costs on decentralized exchanges are independent of order execution, highlighting the significance of economies of scale (lower proportional costs for larger liquidity provision orders) and dynamic liquidity cycles (managing the frequency of fixed cost payments). Strategic brokers in \citet{Cimon2021} provide liquidity alongside exogenous market-makers in a static setting. We complement this approach by modelling network externalities inherent in the coordination problem of heterogeneous liquidity providers. In our dynamic setup, this allows us to pin down the equilibrium duration of liquidity cycles and the relative importance of gas fixed costs.

Our paper relates to a nascent and fast-growing literature on the economics of decentralized exchanges. Many studies \citep[e.g.,][]{aoyagi2020,aoyagiito2021,park2022} focus on the economics of constant-function automated market makers, which do not allow liquidity providers to set price limits. In this restrictive environment, \citet{CapponiJia2021} argue that market makers have little incentives to update their position upon the arrival of news to avoid adverse selection, since pro-rata clearing gives an advantage to arbitrageurs. \citet{LeharParlour2021} solve for the equilibrium pool size in a setting where liquidity providers fully internalize information costs without rushing to withdraw positions at risk of being sniped. We argue that on exchanges that allow for limit or range orders, the cost of actively managing positions becomes a first-order concern, as liquidity providers need to re-set the price limits once posted liquidity no longer earns fees. Our empirical result on economies of scale echoes the argument in \citet{BarbonRanaldo2021}, who compare transaction costs on centralized and decentralized exchanges and find that high gas prices imply that the latter only become competitive for transactions over US\$100,000. \citet{HasbrouckRiveraSaleh2022} argue that liquidity providers require remuneration.  We complement the argument by stating that high fees might be \emph{necessary} for some liquidity providers to cover the fixed costs of managing their position. In line with our theoretical predictions, \citet{caparros2023blockchain} find that liquidity providers reposition their quotes more often on Uniswap V3 pools built on Polygon, which features substantially lower gas fees. Finally, \citet{Heimbach2023} document that after accounting for price impact, concentrated liquidity on Uniswap v3 pools results in increased returns for sophisticated participants but losses for retail traders.

Despite higher gas costs, decentralized exchanges may hold advantages over centralized venues. \citet{han2022trust} demonstrate Uniswap frequently leads price discovery compared to centralized exchanges such as Binance, despite the latter having higher trading volume. \citet{capponi2023price} find that the fee paid by traders to establish execution priority unveils their private information, and therefore contributes to price discovery. \citet{ASPRIS2021101845} argue that decentralized exchanges offer better security than their centralized counterparts since assets are never transferred to the custody of a third party such as an exchange wallet. In turn, \citet{BrolleyZoican2022b} make the point that decentralized exchanges may be able to reduce overall computational costs associated with latency arbitrage races, as they eliminate long-term co-location subscriptions.

Our paper is related to both the finance literature that examines whether make-take fees affect market quality and to the economics literature on two-sided markets and platform competition.  Broadly, our work differs from the finance literature in that we explicitly consider equity markets as markets for liquidity without focusing on the order choice decision, and our work differs from the economics literature in that we explicitly analyze an equity market as a market for liquidity.  The main insight that this brings is that market participants are both large and strategic, compared to  smaller players in consumer-facing markets that are often analyzed in the economics literature.

\section{Model\label{sec:model}}
\paragraph{Asset and agents.} Consider a continuous time model of trade in a single token $\T$ with expected value $v_t>0$. Three risk neutral trader types consummate trade in this market: a continuum of liquidity providers ($\LP$s), liquidity traders ($\LT$s), and arbitrageurs ($\A$). Trade occurs either because public news  triggers a change in the common value of the asset, or because market participants have heterogeneous private values for the asset.

Arrival times of news and private value shocks follow independent Poisson processes with rates $\eta\in\left(0,1\right)$ and $1-\eta$, respectively.\footnote{This is without loss of generality, as what matters in the model is the \emph{relative} arrival rate of news relative to liquidity traders.} For notational compactness, we first characterize the generic shock distribution and then describe its effects on arbitrageurs or liquidity traders. Conditional on an event at time $t$, the asset value changes to $v_t\left(1+{\cal I}\tilde{\delta}\right)$ for all traders in the case of a common value shock, or for an arriving  ($\LT$) in the case of the private value shock.  Here,  ${\cal I}$ is an indicator that takes on the value of $1$ if the taker buys and $-1$ if the taker sells. The value innovation $\tilde{\delta}$ has a probability density
\begin{equation}
    \phi\left(\delta\right)=\frac{1}{2\Delta \sqrt{1+\delta}} \; \; \text{ for } \delta\in\left[0,\Delta^2-1\right],
\end{equation}
thus $\sqrt{1+\tilde{\delta}}$ is uniformly distributed between $\left[1,\Delta\right]$. This assumption is innocuous and made for tractability purposes.

When news arrives, the innovation is to the common value of the token. (As we are agnostic as to the source of value of cryptocurrencies, this common value shock could include the possibility of resale on another exchange.) After such a shock, an arbitrageur $\textbf{A}$ trades with the liquidity providers whenever profitable, and $\LP$s face an adverse selection loss. Conversely, when a liquidity trader enters the market, they experience a private value shock --- and liquidity providers continue to value the token at $v_t$. In what follows for expositional simplicity, as in \citet{foucault2013liquidity}, we focus on a one sided market in which liquidity takers act as buyers, and news lead to an increase in token value.

Liquidity providers ($\LP$) differ in their endowments of the token. Each provider $i$ can supply at most $q_i \diff i$  of the token, where $q_i$ follows an exponential distribution with scale parameter $\lambda$. The right skew of the distribution captures the idea that there are many  low-endowment liquidity providers such as retail traders, but few high-capital $\LP$s such as sophisticated quantitative funds.   Heterogeneity in $\LP$ size is captured by $\lambda$, where a larger $\lambda$ naturally corresponds to a larger dispersion of endowments and larger aggregate liquidity supply. Given the endowment distribution, collectively $\LP$s  supply at most
\begin{equation}
   S  =   \int_0^\infty q_i \frac{1}{\lambda} e^{-\frac{q_i}{\lambda}} \diff i = \lambda
\end{equation}
tokens.

\paragraph{Trading environment.} Traders can interact in two liquidity pools in which token trade occurs against a num\'{e}raire asset (cash). At the start of the trading game, each liquidity provider (LP) deposits liquidity to a single pool within a symmetric price band around the current asset value $\left[\frac{v}{\left(1+r\right)^2}, v(1+r)^2\right]$, where $r \geq 0$. Here, we make use of the fact that V3 features ``price bands,''  and thus liquidity can be consumed with a bounded price impact. Within this range, prices in both pools satisfy a constant product bonding curve as in \citet{Uniswapv3Core2021}. In particular, for pool $k$,
\begin{equation}
    \underbrace{\left(T_k+\frac{L_k}{\sqrt{v}\left(1+r\right)}\right)}_\text{virtual token reserves}\underbrace{\left(T_k v + L_k\frac{\sqrt{v}}{1+r}\right)}_\text{virtual numeraire reserves}=L_k^2,
\end{equation}
where $T_k$ is the amount of tokens deposited on pool $k$ and $L_k$ is the liquidity level of pool $k$, defined as
\begin{equation}
    L_k=\frac{T_k}{\frac{1}{\sqrt{v}}-\frac{1}{\sqrt{v}}\left(1+r\right)}.
\end{equation}
To purchase $\tau$ tokens, a trader needs to deposit an amount $n\left(\tau\right)=\tau T_k \frac{v\left(1+r\right)}{\tau+\left(1+r\right)\left(T_k-\tau\right)}$ of num\'{e}raire into the pool, where $n\left(\tau\right)$ is the solution to the invariance condition
\begin{equation}
    \underbrace{\left(T_k-\tau+\frac{L_k}{\sqrt{v}\left(1+r\right)}\right)}_\text{virtual token reserves}\underbrace{\left(T_k v + n\left(\tau\right) +  L_k\frac{\sqrt{v}}{1+r}\right)}_\text{virtual numeraire reserves}=L_k^2.
\end{equation}

Fees are levied on liquidity takers  as a fraction of the value of the trade and distributed pro rata to liquidity providers.  Crucially, the pools have different fees.  One pool charges a low fee, and one pool charges a high fee which we denote $\ell$ and $h$ respectively.  Specifically, to purchase $\tau$ units of the token on the low fee pool, the total cost to a taker is $\left(1+\ell\right) n\left(\tau, T_\ell\right)$. The $\LP$s  in the pool receive $\ell n\left(\tau, T_\ell\right)$ in fees. In addition, consistent with gas costs on Ethereum, all traders incur a fixed execution cost $\Gamma \diff i >0$ to interact with the market.

Figure \ref{fig:timing_revised} illustrates the timing of the model.

\begin{figure}[H]
    \centering
\begin{tikzpicture}
    \draw[|->, thick] (0,0) -- (15.75,0);
    \draw[|->, thick] (0,1.5) -- (15.75,1.5);

    \node at (15.75,0.5) {Pool H};
    \node at (15.75, 2) {Pool L};

    \node[align=left] at (1.5, 0.75) {$\LP$s deposit $q_i$ \\ in pool $k\in\left\{L,H\right\}$};

    \draw[->,blue] (1,2)--(1,1.5);
    \node[blue, align=center] at (1,2.4) {Small \\ LT};

    \draw[->,blue] (2.5,2)--(2.5,1.5);
    \node[blue, align=center] at (2.5,2.4) {Small \\ LT};

    \draw[->,blue,thick] (4,2)--(4,1.5);
    \draw[->,blue,thick] (4,1.5)--(4,0);
    \node[blue, align=center] at (4,2.4) {\bfseries Large \\ \bfseries LT};

    \draw[dotted,thick,red] (4.5,1.5) -- (4.5, 0);
    \node[red, align=left] at (5.25,0.75) {News \\ $\tilde{\delta}\in\left(\ell,h\right)$}; 

    \draw[->,red] (6,2)--(6,1.5);
    \node[red, align=left] at (6,2.4) {$\A$ trades \\ on pool $L$};

    \draw[->,thick](8,2)--(8,1.5);
    \node[align=center] at (8,2.4) {$\LP$ rebalance \\ on pool $L$};

    \draw[->,blue,thick] (9.75,2)--(9.75,1.5);
    \draw[->,blue,thick] (9.75,1.5)--(9.75,0);
    \node[blue, align=center] at (9.75,2.4) {\bfseries Large \\ \bfseries LT};

    \draw[->,blue] (10.75,2)--(10.75,1.5);
    \node[blue, align=center] at (10.75,2.4) {Small \\ LT};

    \draw[dotted,thick,red] (11.25,1.5) -- (11.25, 0);
    \node[red, align=center] at (11.75,0.75) {News \\ $\tilde{\delta}>h$}; 

    \draw[->,red] (12.5,2)--(12.5,1.5);
    \draw[->,red] (12.5,1.5)--(12.5,0);
    \node[red, align=left] at (12.5,2.6) {\textbf{A} trades \\ on both \\ pools};

    \draw[->,thick](14.75,2)--(14.75,1.5);
    \draw[->,thick](14.75,1.5)--(14.75,0);
    \node[align=left] at (14.75,2.6) {$\LP$ rebalance \\ on both \\ pools};

\end{tikzpicture}
    \caption{Model timing}
    \label{fig:timing_revised}
\end{figure}

To ensure the possibility of liquidity re-balancing in both pools, we assume that innovations are large enough to ensure that $\LP$s may need to rebalance their position on the high fee pool or:

\begin{ass}\label{ass:Delta} The size of innovations are sufficiently large so that there is a positive probability that liquidity providers need to re-balance on the high-fee pool. That is, $\Delta>\left(1+r\right)\sqrt{1+h}$.
\end{ass}

\subsection{Equilibrium}

\subsubsection{Optimal trade size}

First, consider the decisions of arbitrageurs and liquidity traders holding a value $v\left(1+\delta\right)$ for the asset. Faced with pool sizes of $T_\ell$ and $T_h$ in the low and high pool respectively, their optimal trade on pool $k$ maximizes their expected profit, net of fees and price impact:
\begin{equation}
    \max_{\tau} \text{Profit $\LT$}\left(\tau,\delta\right) \equiv \tau v \left(1+\delta\right) - \left(1+f_k\right) \tau T_k \frac{v\left(1+r\right)}{\tau+\left(1+r\right)\left(T_k-\tau\right)},
\end{equation}
which yields the optimal trade quantity:
\begin{equation}\label{eq:optimal_trade_size}
    \tau^\star\left(\delta\right)=T_k\min\left\{1,\frac{1+r}{r}\max\left\{0,1-\sqrt{\frac{1+f_k}{1+\delta}}\right\}\right\}.
\end{equation}
From equation \eqref{eq:optimal_trade_size}, a trader with valuation $v\left(1+\delta\right)$ only trades on pool $k$ if the gains from trade are larger than the liquidity fee, i.e., $\delta>f_k$. Further, if $\delta>\left(1+f_k\right)\left(1+r\right)^2-1$ so that the gains from trade are larger than the maximum price impact, then the trader consumes all available liquidity in the pool.

\subsubsection{Fee revenue for liquidity providers from private value trades}

The revenue for liquidity providers can be expressed as the product of the pool fee and the num\'{e}raire deposit required from liquidity traders to purchase \(\tau^\star\) token units, denoted by \(n(\tau^\star, T_k)\). That is, 
\begin{equation}\label{eq:revenue_raw}
    f_k n\left(\tau^\star\left(\delta\right), T_k\right) = f_k v T_k \min\left\{1+r,\frac{1+r}{r}\max\left\{0,\sqrt{\frac{1+\delta}{1+f_k}}-1\right\}\right\}.
\end{equation}

If the innovation $\delta$ corresponds to a private rather than common value shock, then an arbitrageur optimally steps in to reverse the liquidity trade as described in \citet{LeharParlour2021}. In this case, liquidity providers effectively earn double the fee revenue in \eqref{eq:revenue_raw} without affecting the capital structure of the pool; there is neither a capital gain nor a loss for the $\LP$s. 

The fee revenue in \eqref{eq:revenue_raw} scales linearly with the size of the pool $T_k$. Since fee proceeds are distributed pro-rata among liquidity providers based on their share $\frac{q_i}{T_k}$, it follows that fee revenue an $\LP$ with endowment $q_i$ providing liquidity on pool $k$  increases linearly in their endowment:
\begin{align}\label{eq:fee_revenue_delta}
    \text{FeeRevenue}_{i,k}\left(\delta\right)&=2\frac{q_i}{T_k}f_k n\left(\tau^\star, T_k\right) \nonumber \\
    &= 2q_i v f_k \min\left\{1+r,\frac{1+r}{r}\max\left\{0,\sqrt{\frac{1+\delta}{1+f_k}}-1\right\}\right\}.
\end{align}

The expression in \eqref{eq:fee_revenue_delta} denotes the fee revenue conditional on the private value $\delta$ of the incoming trade. To compute the expected fee revenue, we integrate this expression across all posible value shocks:
\begin{align}
\mathbb{E}\text{FeeRevenue}_{i,k}&=\int_{\delta=1}^{\Delta^2-1} \text{FeeRevenue}_{i,k}\left(\delta\right) \phi\left(\delta\right)\diff \delta \nonumber \\
&=q_i \underbrace{v \frac{f_k (r+1) \left(2 \Delta -r\sqrt{f_k+1} -2 \sqrt{f_k+1}\right)}{\Delta }}_{\equiv \mathcal{L}\left(f_k\right)},
\end{align}
where we define $\mathcal{L}\left(f_k\right)$ as the \emph{liquidity yield}: that is, the per-unit $\LP$ fee revenue from supplying liquidity to $\LT$s in pool $k$.

\begin{lem}\label{lem:liq_revenue} There exists a threshold fee level $\overline{f}>0$ such that the liquidity revenue $\mathcal{L}\left(f_k\right)$ first increases in the pool fee $f_k$ for $f\leq\overline{f}$, then decreases in the pool fee for $f>\overline{f}$.
\end{lem}

Lemma \ref{lem:liq_revenue} points out to a non-linear relationship between fee levels and liquidity yield. Initially, as fees increase, the enhanced revenue from higher fees outweighs the decrease in trading volume due to increased transaction costs, resulting in a net gain in revenue. However, beyond a certain fee threshold, the drop in trading volume dominates the larger fee, leading to a decrease in overall revenue. A salient implication is that if pool fees are large enough, the liquidity yield on the high fee pool may exceed the yield on the low-fee pool.

\subsubsection{Adverse selection cost for liquidity providers}

If news occurs (i.e., if $\delta$ represents a common value shock), liquidity providers trade against arbitrageurs rather than liquidity traders. In this case, there is no subsequent price reversal following the initial trade. The capital structure of the liquidity pool changes, as arbitrageurs remove the more valuable asset: i.e., buy tokens upon a positive common value shock. While $\LP$s earn fee revenues on arbitrage trades, they also incur adverse selection losses by trading against the direction of the news. Moreover, if the magnitude of news is large enough that arbitrageurs remove all tokens supplied in the price range, then $\LP$s face additional costs, that is a gas fee $\Gamma \diff i$ to re-balance liquidity around the new asset value.

Table \ref{tab:fee_rev} delineates the $\LP$ fee revenue from selling tokens to arbitrageurs, as well as the marked-to-market value of the tokens sold. If the size of news ($\delta$) does not exceed the pool fee, then arbitrageurs do not trade since  the potential profit does not justify the transaction cost. Conversely, if the news size is larger than pool fee, then arbitrageurs execute a trade proportional to the size of the pool, and they exhaust the available liquidity on the price range if the news is large enough: specifically, if $\delta>\left(1+f_k\right)\left(1+r\right)^2-1$. The profit for liquidity providers in each scenario is the difference between the revenue and the marked-to-market value. Notably, the profit is consistently negative, since $\LP$s are trading against the direction of news.

\begin{table}[]
    \caption{Fee revenue and capital losses on arbitrage trades}
    \label{tab:fee_rev}
\begin{center}
    \begin{tabular}{lll}
\toprule
News size & Revenue (numeraire) & Marked-to-market token value \\ 
\cmidrule{1-3}
$\delta\leq f_k$     & 0 & 0  \\
$\delta \in \left(f_k, \left(1+f_k\right)\left(1+r\right)^2-1\right]$     &  $v q_i \frac{1+r}{r}\left(1+f_k\right)\left(\sqrt{\frac{1+\delta}{1+f_k}}-1\right)$ & $v q_i \frac{1+r}{r}\left(1+\delta\right)\left(1-\sqrt{\frac{1+f_k}{1+\delta}}\right)$ \\
$\delta>\left(1+f_k\right)\left(1+r\right)^2-1$ & $vq_i\left(1+f_k\right)\left(1+r\right)$ & $v q_i \left(1+\delta\right)$ \\
\bottomrule
\end{tabular}
\end{center}

\end{table}

The expected $\LP$ profit from trading with arbitrageurs equals $-q_i v \times \mathcal{A}\left(f_k\right)$,
where $\mathcal{A}\left(f_k\right)$ is the per-unit adverse selection cost from liquidity provision in pool $k$:
\begin{align}\label{eq:AScost}
\mathcal{A}\left(f_k\right)&=\mathbb{P}\left(f_k<\delta \leq (1+f_k)(1+r)^2-1\right) \times \frac{1+r}{r}\mathbb{E} \left[\left(1+f_k\right)+\left(1+\delta\right)-2\sqrt{\left(1+\delta\right)\left(1+f_k\right)}\right] + \nonumber \\
    &+ \mathbb{P}\left(\delta>(1+f_k)(1+r)^2-1\right) \times \left[\mathbb{E}\left(1+\delta\right)\right]-\left(1+f_k\right)\left(1+r\right) \Big\}.
\end{align}
 
\begin{lem}\label{lem:advsel_fees} The adverse selection cost $\mathcal{A}\left(f_k\right)$ decreases in the pool fee $f_k$. In particular, the high-fee pool has a lower adverse selection cost than the low-fee pool. 
\end{lem}

\begin{figure}[H]
\caption{\label{fig:liqrev_as} Liquidity yield and adverse selection cost}
\begin{minipage}[t]{1\columnwidth}%
\footnotesize
This figure illustrates the expected fee yield from liquidity trades (left panel) and the adverse selection cost (right panel), as a function of the pool fee $f$. Parameter values: $r=0.001$, $\lambda=1$, $\eta=0.1$, and $\Delta=1.1\left(1+r\right)\sqrt{1+h}$.
\end{minipage}

\vspace{0.05in}

\begin{centering}
\includegraphics[width=\textwidth]{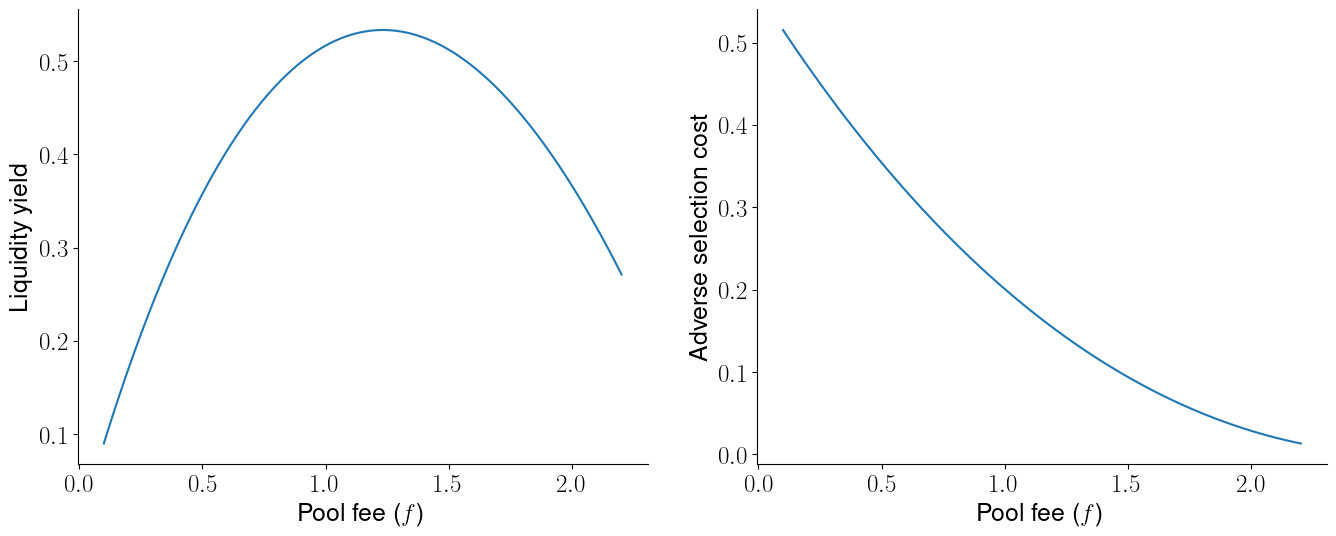}
\par\end{centering}

\end{figure}

Lemma \ref{lem:advsel_fees} indicates that higher pool fees lower adverse selection costs through two mechanisms: First, they increase compensation per unit traded for liquidity providers (\(\LP\)s), enhancing their returns on trades with arbitrageurs. Second, higher fees discourage arbitrageur activity, effectively reducing the volume of informed trades. Figure \ref{fig:liqrev_as} showcases the results in Lemmas \ref{lem:liq_revenue} and \ref{lem:advsel_fees} and illustrates the comparative statics of liquidity yield and adverse selection cost with respect to the pool fee.

Liquidity rebalancing costs arise only when news events are large enough to deplete all available liquidity within a given price range, pushing liquidity providers' (\(\LP\)s) positions ``out of range.'' Rebalancing only occurs post-news, since equally large liquidity trades would be reversed by arbitrageurs. Conditional on news arrival, the expected cost of rebalancing is 
\begin{equation}
    \mathcal{C}\left(k\right)= \mathbb{P}\left(\delta>(1+f_k)(1+r)^2-1\right) \Gamma = \Gamma\left(1-\frac{\sqrt{1+f_k}\left(1+r\right)}{\Delta}\right),
\end{equation}
which is decreasing in the pool fee $f_k$. This result is straightforward: smaller news events can cause arbitrageurs to deplete liquidity in low-fee pools, whereas it takes larger news to do the same in high-fee pools. Consequently, \(\LP\)s in lower fee pools face more frequent rebalancing and incur higher fixed costs.

\subsubsection{Liquidity provider pool choice}

Liquidity providers face a choice between the low and high fee pool or not participating in the market. An $\LP$ of size $q_i$ earns expected profit
\begin{align}\label{eq:profitLP}
    \pi_L &= q_i \left[\left(1-\eta\right) \mathcal{L}\left(\ell\right)-\eta \mathcal{A}\left(\ell\right)\right]-\eta \Gamma\left(1-\frac{\sqrt{1+\ell}\left(1+r\right)}{\Delta}\right) \text{ and } \\
    \pi_H &= q_i \left[\left(1-\eta\right) \mathcal{L}\left(h\right)-\eta \mathcal{A}\left(h\right)\right]-\eta \Gamma\left(1-\frac{\sqrt{1+h}\left(1+r\right)}{\Delta}\right) \nonumber,
\end{align}
from choosing pool $L$ or $H$, respectively. Equation \eqref{eq:profitLP} underscores the trade-off faced by liquidity providers (\(\LP\)s): balancing the liquidity yield from trades with liquidity traders (\(\LT\)s) against the adverse selection costs and the fixed gas costs associated with re-balancing their position.

First, consider the choice of participating in the market. An agent  only provides liquidity on pool $k$ if she is able to break even -- that is, if her endowment $q_i$ is large enough. We define $\underline{q}_L$ and $\underline{q}_H$ as the thresholds at which the participation constraints $\pi_L\left(q\right)=0$ and $\pi_H\left(q\right)=0$ are satisfied, respectively. If \(\underline{q}_k \geq 0\) for a pool \(k\), it indicates that any \(\LP\) with an endowment \(q_i\) at least equal to \(\underline{q}_k\) can join pool \(k\) and expect to earn a positive profit, with the marginal entrant breaking even.  Conversely, if \(\underline{q}_k < 0\), it suggests that pool \(k\) is not economically viable as the participation constraint is breached for all \(\LP\)s.

\begin{ass}\label{ass:eta}
    To avoid trivial cases, we focus on the case that both markets are potentially viable, or equivalently the intensity of news is low enough:
    \begin{equation}
        \eta \leq \min_k \frac{\mathcal{L}\left(k\right)}{\mathcal{L}\left(k\right)+\mathcal{A}\left(k\right)},
    \end{equation}
    such that $\underline{q}_k\geq 0$.
\end{ass}

Next, consider the choice between pools. Liquidity provider $i$ chooses the low-fee pool if and only if
\begin{equation}\label{eq:pi_diff}
    \pi_L-\pi_H = q_i \left[\left(1-\eta\right)\left(\mathcal{L}\left(\ell\right)-\mathcal{L}\left(h\right)\right)+\eta \underbrace{\left(\mathcal{A}\left(h\right)-\mathcal{A}\left(\ell\right)\right)}_{<0} \right] - \underbrace{\Gamma \frac{\eta(1+r)}{\Delta}\left(\sqrt{1+h}-\sqrt{1+\ell}\right)}_{>0}>0.
\end{equation}
Liquidity providers in the high-fee pool face both lower adverse selection and rebalancing costs. Therefore, the low-fee pool can only be chosen in equilibrium if it offers a higher liquidity yield, specifically if \(\mathcal{L}(\ell) - \mathcal{L}(h) > 0\), and if the intensity of news \(\eta\) is sufficiently low. Otherwise, all liquidity providers prefer to supply tokens to the high-fee pool if the participation constraint is satisfied. Further, equation \eqref{eq:pi_diff} highlights the economies of scale embedded in liquidity provision with fixed rebalancing costs. That is, if a liquidity provider of size $q$ prefers the low fee pool, then any liquidity provider with a larger endowment,  $\widetilde{q}>q$, also prefers the low fee pool.


Proposition \ref{prop:equilibria} characterizes the equilibrium liquidity provision.

\begin{prop}\label{prop:equilibria}
\begin{enumerate}
    \item [i.]
If $\eta>\frac{\mathcal{L}\left(l\right)-\mathcal{L}\left(h\right)}{\mathcal{L}\left(l\right)-\mathcal{L}\left(h\right)+\mathcal{A}\left(l\right)-\mathcal{A}\left(h\right)}$, then  all $\LP$s with $q_i>\underline{q}_h$ deposit liquidity on the high fee pool. 
\item [ii.] Otherwise, there exists a unique fragmented equilibrium characterized by marginal trader $\qmg^\star>\underline{q}_h$ which solves
\begin{equation}\label{eq:mg_eq}
    \qmg^\star = \Gamma \frac{\eta(1+r\left(\sqrt{1+h}-\sqrt{1+\ell}\right))}{\Delta \left[\left(1-\eta\right)\left(\mathcal{L}\left(\ell\right)-\mathcal{L}\left(h\right)\right)+\eta \left(\mathcal{A}\left(h\right)-\mathcal{A}\left(\ell\right)\right) \right]}
\end{equation}
such that all $\LP$s with $q_i\in\left(\underline{q}_h,\qmg^\star\right]$ deposit liquidity in the high fee pool and all $\LP$s with $q_i>\qmg^\star$ choose the low fee pool.
\end{enumerate}
\end{prop}

\begin{figure}[H]
\caption{\label{fig:region_equilibrium} Fragmented and single-pool equilibria}
\begin{minipage}[t]{1\columnwidth}%
\footnotesize
This figure plots the existence conditions for a fragmented market equilibrium, as described in Proposition \ref{prop:equilibria}, for various values of the news intensity ($\eta$) on the y-axis and liquidity fee on pool $L$ on the x-axis. Parameter values: $r=0.001$, $h=2$, $\lambda=1$, $\eta=0.1$, and $\Delta=1.1\left(1+r\right)\sqrt{1+h}$.
\end{minipage}

\vspace{0.05in}

\begin{centering}
\includegraphics[width=\textwidth]{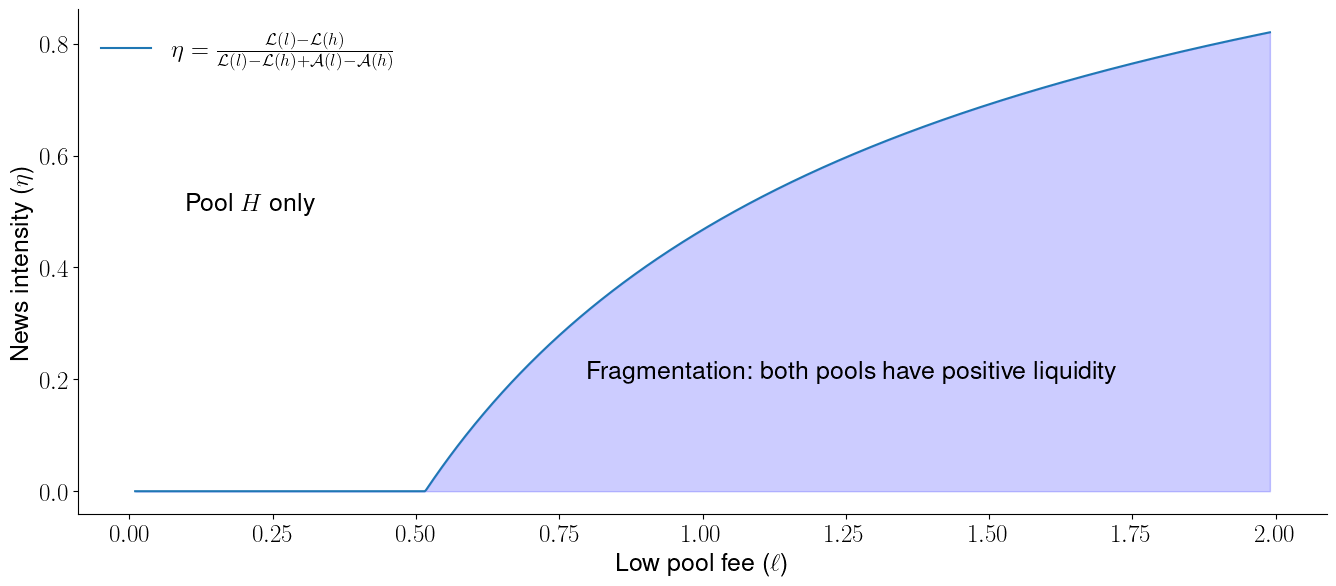}
\par\end{centering}

\end{figure}

Figure \ref{fig:region_equilibrium} illustrates the equilibrium regions in Proposition \ref{prop:equilibria}. When news intensity $\eta$ is high, or pool $H$ offers a substantially higher fee than pool $L$, liquidity suppliers gravitate towards pool $H$, resulting in a single-maker equilibrium. Conversely, a lower $\eta$ translates to lower adverse selection costs. If this is the case, or if the fee differential between the two pools is low, liquidity providers with large endowments $q$ migrate to the lower-fee pool to compete for order flow from small traders, causing liquidity to fragment between the two pools.

Proposition \ref{cor:comp_stat_ms} establishes the impact of gas prices on the two pools' liquidity market shares. We can compute the liquidity market share of the low-fee pool in a fragmented equilibrium as
\begin{equation}
w_\ell=\frac{\exp\left(-\frac{\qmg-\underline{q}_h}{\lambda}\right)\left(\qmg+\lambda\right)}{\underline{q}_h+\lambda} \leq 1,
\end{equation}
with equality for $\Gamma=0$. That is, as fixed costs drop to zero, the low fee pool asymptotically captures the full market share.

\begin{prop}\label{cor:comp_stat_ms}
In equilibrium, the market share of the low fee pool $w_\ell$ decreases in the gas cost ($\Gamma$). 

\end{prop}

We stress the critical role of fixed gas costs in driving market fragmentation. Since liquidity fee revenues and adverse selection costs are distributed pro-rata, in the absence of gas fees, all liquidity providers (\(\LP\)s) would converge on a single pool --- the one offering the optimal balance between fee yield and informational costs. For instance, if \(\Gamma=0\), all \(\LP\)s would select pool \(L\) if the news arrival rate is sufficiently low, as defined by \(\eta\leq\frac{\mathcal{L}(l)-\mathcal{L}(h)}{\mathcal{L}(l)-\mathcal{L}(h)+\mathcal{A}(l)-\mathcal{A}(h)}\), or choose pool \(H\) otherwise. It is the introduction of fixed costs that drives \(\LP\)s to segregate into different pools based on their size.

\begin{figure}[H]
\caption{\label{fig:liqshares} Liquidity shares and gas costs}
\begin{minipage}[t]{1\columnwidth}%
\footnotesize
This figure illustrates the equilibrium liquidity market shares (left panel) and the aggregate liquidity supply on the two pools (right panel), as a function of the gas fee $\Gamma$. Parameter values: $r=0.001$, $h=2$, $\ell=1$, $\lambda=1$, $\eta=0.1$, and $\Delta=1.1\left(1+r\right)\sqrt{1+h}$.
\end{minipage}

\vspace{0.05in}

\begin{centering}
\includegraphics[width=\textwidth]{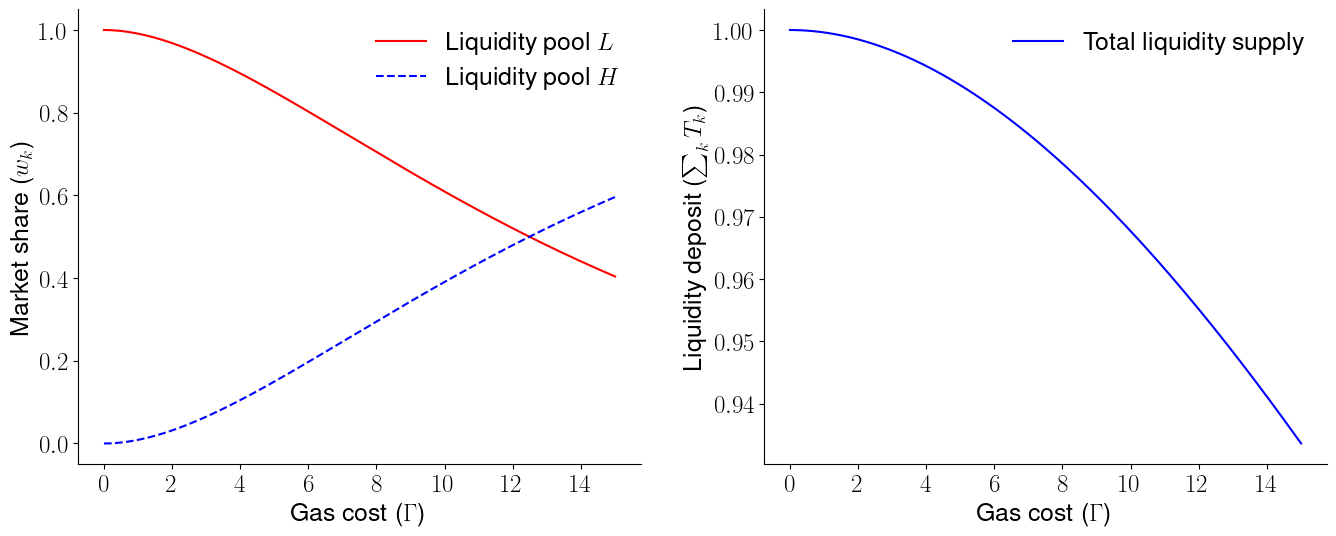}
\par\end{centering}

\end{figure}

Figure \ref{fig:liqshares} shows that the market share of the low fee pool decreases in the gas cost $\Gamma$. A larger gas price increases the costs of re-balancing upon the arrival of large enough news, everything else equal, and incentivizes smaller $\LP$s to switch from the low fee pool to the high fee pool, since the arbitrageurs are less likely to fully consume liquidity there. Further, the right panel illustrates the extensive margin effect of gas prices: any increase in gas costs leads to a decrease in aggregate liquidity supply as some $\LP$ with low endowments are driven out of the market (that is, the threshold $\underline{q}_h$ increases in $\Gamma$).

\subsection{Pool fragmentation and market quality}

We measure market quality by the realized gains from trade of liquidity traders. If the asset is traded on a sequence of pools, where $f_k$ and $T_k$ represent the fees and liquidity deposits on pool $k$, respectively, the expected gains from trade for liquidity traders are
\begin{equation}\label{eq:gains}
    \text{GainsFromTrade}\left(\left\{f_k\right\}_k\right)= v \mathbb{E}\left[ \sum_k  \tau^\star\left(f_k,\delta\right) \times \delta\right],
\end{equation}
where $\tau^\star\left(\delta\right)=T_k\min\left\{1,\frac{1+r}{r}\max\left\{0,1-\sqrt{\frac{1+f_k}{1+\delta}}\right\}\right\}$ is the optimal \textbf{LT} trade size, as defined in equation \eqref{eq:optimal_trade_size}.

Suppose an asset is traded on a single pool that imposes a liquidity fee $f$. From equation \eqref{eq:gains} it follows that the gains from trade for an $\LT$ with private value $1+\delta$ are equal to
\begin{equation}
        \text{GainsFromTrade}\left(f\mid \delta\right)=v \delta T_f\min\left\{1,\frac{1+r}{r}\max\left\{0,1-\sqrt{\frac{1+f}{1+\delta}}\right\}\right\}.
\end{equation}
The total token supply on the single pool equals $T_f=e^{-\underline{q}_f \lambda}\left(\underline{q}_f+\lambda\right)$, where $\underline{q}_f$ is the marginal liquidity provider such that all $\LP$s with endowment $q_i>\underline{q}_f$ join the pool. Here, the magnitude of the liquidity fee drives the trade-off between the participation of liquidity providers (\(\LP\)) and trading costs. A lower fee \( f \) results in fewer \(\LP\)s offering liquidity, a lower token supply $T_f$, which limits gains from trade for liquidity traders. In contrast, a higher fee increases trading costs, potentially outweighing the benefits of increased \(\LP\) participation.

\begin{prop}\label{prop:optimality}
For any single-pool fee $f\geq 0$, there exists a set of fees \( \{h, \ell\} \) for a two-pool fragmented market, where \( h = f \) and \( h > \ell \), that guarantees equal or higher gains from trade in a fragmented market compared to the single-pool market.
\end{prop}

Proposition \ref{prop:optimality} suggests that fragmentation with multiple fee levels improves market quality. Specifically, it is always possible to devise a fee structure in a fragmented market that yields (weakly) higher gains from trade than a single-fee market. The logic is as follows: First, the highest fee in the fragmented market is set equal to the single pool fee, ensuring that the marginal $\LP$ participating the market is the same across both scenarios (i.e., the $\LP$ with endowment $\underline{q}_h$). This condition guarantees the same aggregate liquidity supply in fragmented and non-fragmented markets. Second, a lower fee is then chosen for another pool to attract liquidity providers with higher token endowments, resulting in larger trade sizes per unit of supplied liquidity. This combination of larger liquidity trades and unchanged aggregate liquidity supply leads to higher gains from trade in a fragmented market.

\begin{figure}[H]
\caption{\label{fig:gft} Gains from trade and market structure}
\begin{minipage}[t]{1\columnwidth}%
\footnotesize
The figure plots the expected gains from trade across $\LT$s,
$$\text{GainsFromTrade}=\int_0^{\Delta^2-1} v\delta\tau^\star\left(\delta\right)\phi\left(\delta\right)\diff \delta,$$
on both a single pool with a high fee as well as on fragmented pools, as a function of the gas cost $\Gamma$. Parameter values: $r=0.001$, $h=2$, $\ell=1$, $\lambda=1$, $\eta=0.1$, and $\Delta=1.1\left(1+r\right)\sqrt{1+h}$.
\end{minipage}

\vspace{0.05in}

\begin{centering}

\includegraphics[width=\textwidth]{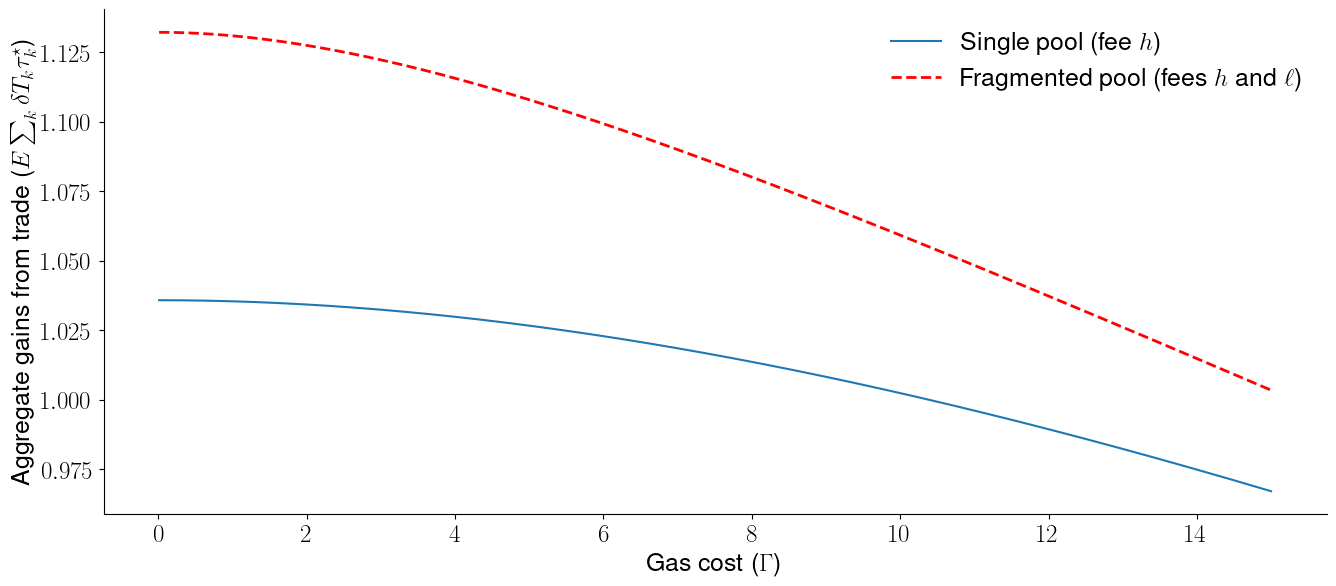}

\par\end{centering}

\end{figure}

Figure \ref{fig:gft} illustrates the result. As gas price increase, the gains from trade drop in both single-pool and fragmented markets, primarily because more $\LP$s are priced out which results in lower liquidity supply and higher price impact. Nevertheless, irrespective of the level of gas costs, the gains from trade are higher in the fragmented market.

We note that the argument discussed in this section is valid for \emph{any} single-fee pool, including an optimally designed one. In essence, if a fragmented fee structure can be designed to achieve higher gains from trade compared to an arbitrary single-pool fee, then a fee structure that dominates the optimally set single-pool fee achieves higher gains from trade than any single-fee pool.

\subsection{Model implications and empirical predictions}

\begin{pred}\label{pred:comp_stat_Gamma}
The liquidity market share of the low-fee pool  decreases in the gas fee $\Gamma$.
\end{pred}

Prediction \ref{pred:comp_stat_Gamma} follows directly from Proposition \ref{cor:comp_stat_ms} and Figure \ref{fig:liqshares}. A higher gas price increases the cost of liquidity re-balancing. Given that re-balancing is more frequently required in the low-fee pool due to more intense arbitrage activity, liquidity providers, particularly those with smaller endowments, optimally migrate to the high-fee pool in response to a gas cost increase.

\begin{pred}\label{pred:clienteles}
$\LP$s on the low-fee pool make larger liquidity deposits than $\LP$s on the high-fee pool.
\end{pred}

\begin{figure}[H]
\caption{\label{fig:theory_liqsupply} Liquidity supply on fragmented markets}
\begin{minipage}[t]{1\columnwidth}%
\footnotesize
This figure illustrates the endowment distribution of $\LP$s and their choice of pools in a fragmented market. First, liquidity providers to the left of $\underline{q}_h$ do not provide liquidity on either pool. Next, $\LP$s to the left (right) of the marginal trader $\qmg^\star$ provide liquidity on pool $H$ (pool $L$, respectively). $r=0.001$, $h=2$, $\ell=1$, $\lambda=1$, $\eta=0.1$, $\Gamma=20$, and $\Delta=1.1\left(1+r\right)\sqrt{1+h}$.
\end{minipage}

\vspace{0.05in}

\begin{centering}
\includegraphics[width=\textwidth]{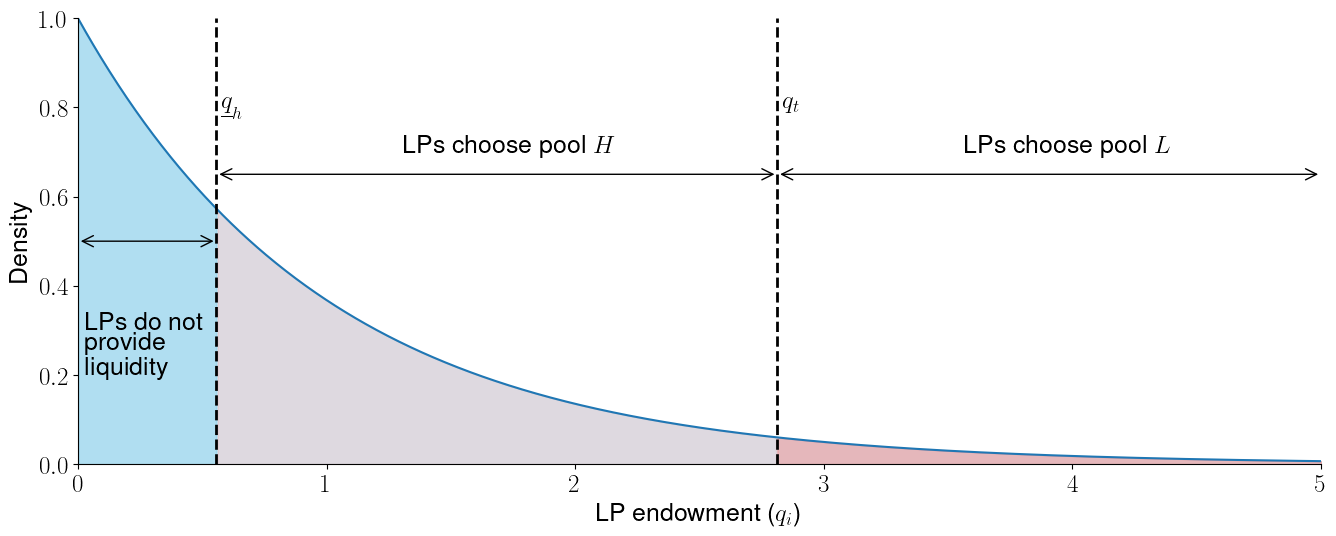}
\par\end{centering}

\end{figure}

Prediction \ref{pred:clienteles} follows from the equilibrium discussion in Proposition \ref{prop:equilibria}. Liquidity providers with large token endowments ($q_i>\qmg$) deposit them in the low-fee pool since they are better positioned to actively manage liquidity due to economies of scale. $\LP$s with lower endowments ($q_i\leq\qmg$) either stay out of the market or choose pool $H$ which allows them to offer liquidity in a more passive manner. Figure \ref{fig:theory_liqsupply} illustrates this prediction by overlaying optimal pool choices on the distribution of $\LP$ endowments. Low-endowment $\LP$s (in blue) that are being rationed out of the market due to high gas cost, medium-endowment $\LP$s (gray) that deposit liquidity on pool $H$, and high-endowment $\LP$s (red) that choose the low-fee pool $L$. 

\begin{pred}\label{pred:trade_size_volume}
The average trade size is higher on pool $H$ than on pool $L$. At the same time, trading volume is higher on pool $L$ than on pool $H$.
\end{pred}

Next, Prediction \ref{pred:trade_size_volume} deals with differences between incoming trades on the two liquidity pools. If liquidity traders and arbitrageurs find it optimal to trade on pool $H$ since $\delta>h$, then they would also trade on pool $L$ since $h>\ell$ and therefore $\delta>\ell$. However, the opposite is not true: $\LT$s and arbitrageurs with $\delta\in\left[\ell,h\right)$ only trade on pool $L$. In equilibrium, only a fraction of traders with sufficiently high private values are drawn to pool $H$.

\begin{pred}\label{pred:ly}
    In a fragmented market equilibrium, the liquidity yield is higher on the low fee pool than on the high fee pool.
\end{pred}

Prediction \ref{pred:ly} is a consequence of Proposition \ref{prop:equilibria}. The high fee pool offers better protection against adverse selection and re-balancing costs. If the low fee pool attracts a positive market share, then it necessarily compensates with a higher liquidity yield.

\begin{pred}\label{pred:clienteles_cs}
The average liquidity deposit on both the low- and- high fee pool increases with gas costs.
\end{pred}

An increase in the gas cost $\Gamma$ has two effects: first, the $\LP$s with the lowest endowments on pool $L$ switch to pool $H$. As a result, the average deposit on pool $L$ increases. Second, the $\LP$s with low endowments on pool $H$ may leave the market. Both channels translate to a higher average deposit on pool $H$, which experiences an inflow (outflow) of relatively high (low) endowment $LP$ following an increase in gas costs.

\begin{pred}\label{pred:updates}
$\LP$s re-balance liquidity more frequently on the low-fee than on the high-fee pool.
\end{pred}

Liquidity providers re-balance their positions in a pool charging a fee $f$ only when the magnitude of news exceeds a threshold, specifically if $\delta > (1+f)(1+r)^2 - 1$. The likelihood of re-balancing given the news is $(1-\frac{\sqrt{1+f}(1+r)}{\Delta})$. Consequently, the duration of a liquidity cycle, expressed as $\frac{1}{\eta(1-\frac{\sqrt{1+f}(1+r)}{\Delta})}$, increases in the pool fee level.

\begin{pred}\label{pred:as}
    Adverse selection cost is higher on the low fee pool than on the high fee pool.
\end{pred}

This prediction follows directly from Lemma \ref{lem:advsel_fees}: a higher pool fee serves as a deterrent to arbitrageurs, particularly if the size of news remains below a threshold.


\section{Data and descriptive statistics \label{sec:data}}
\subsection{Sample construction \label{sec:sample}}

We obtain data from the Uniswap V3 Subgraph, covering all trades, liquidity deposits (referred to as ``mints''), and liquidity withdrawals (referred to as ``burns'') on 4,069 Uniswap v3 pools. The data spans from the protocol's launch on May 4, 2021, up until July 15, 2023. Each entry in our data includes a transaction hash that uniquely identifies each trade and liquidity update on the Ethereum blockchain. Additionally, it provides details such as trade price, direction, and quantity, along with quantities and price ranges for each liquidity update. Moreover, the data also includes wallet addresses associated with initiating each transaction, akin to anonymous trader IDs. The Subgraph data we obtained also provides USD-denominated values for each trade and liquidity mint. We further collect daily pool snapshots from the Uniswap V3 Subgraph, including the end-of-day pool size in Ether and US Dollars, and summary price information (e.g., open, high, low, and closing prices for each pool).

To enhance our dataset, we combine the Subgraph data with public Ethereum data available on \href{https://cloud.google.com/blog/products/data-analytics/ethereum-bigquery-public-dataset-smart-contract-analytics}{Google Big Query} to obtain the position of each transaction in its block, as well as the gas price limit set by the trader and the amount of gas used. Finally, we obtain block-by-block liquidity snapshot data across multiple price ranges from \href{https://www.kaiko.com/}{Kaiko}. 



There are no restrictions to list a token pair on Uniswap. Some pools might therefore be used for experiments, or they might include untrustworthy tokens. Following \citet{LeharParlour2021}, we remove pools that are either very small or that are not attracting an economically meaningful trading volume. We retain liquidity pools that fulfill the following four criteria: (i) have at least one interaction in more than 100 days in the sample, (ii) have more than 500 liquidity interactions throughout the sample, (iii) have an average daily liquidity balance in excess of US\$100,000, and (iv) capture more than 1\% of trading volume for a particular asset pair. 
We exclude burn events with zero liquidity withdrawal in both base and quote assets, as traders use them solely to collect fees without altering their liquidity position.

These basic screens give us a baseline sample of 274 liquidity pools covering 242 asset pairs, with combined daily dollar volume of \$1.12 billion and total value locked (i.e., aggregate liquidity supply) of \$2.53 billion as of July 15, 2023. We capture 24,202,803 interactions with liquidity pool smart contracts (accounting for 86.04\% of the entire universe of trades and liquidity updates). Trading and liquidity provision on Uniswap is heavily concentrated: the five largest pairs (USDC-WETH, WETH-USDT, USDC-USDT, WBTC-WETH, and DAI-USDC) account on average for 86\% of trading volume and 63\% of supplied liquidity.\footnote{WETH and WBTC stand for ``wrapped'' Bitcoin and Ether. Plain vanilla Bitcon and Ether are not compliant with the ERC-20 standard for tokens, and therefore cannot be directly used on decentralized exchanges' smart contracts. USDC (USD Coin), USDT (Tether), and DAI are stablecoins meant to closely track the US dollar.}

\subsection{Liquidity fragmentation patterns}

For 32 out of the 242 asset pairs in our baseline sample, liquidity supply is fragmented across two pools with different fees -- either with 1 and 5 bps fees (5 pairs), 5 and 30 bps fees (6 pairs), or 30 and 100 bps fees (21 pairs).\footnote{In some cases, more than two pools are created for a pair -- e.g., for USDC-WETH there are four pools with 1, 5, 30, and 100 bps liquidity fees. In all cases however, two pools heavily dominate the others: As described in Section \ref{sec:sample} we filter out small pools with less than 1\% volume share or less than \$100,000 liquidity deposits.} Despite being fewer in number, fragmented pairs are economically important: they account on average for 95\% of the capital committed to Uniswap v3 and for 93\% of its dollar trading volume. All major token pairs such as WETH-USDC, WETH-USDT, or WBTC-WETH trade on fragmented pools.

For each fragmented liquidity pair, we label the \emph{low} and the \emph{high} fee liquidity pool to facilitate analysis across assets. For example, the low and high liquidity fees for USDC-WETH are 5 and 30 bps, respectively, but only 1 and 5 bps for a lower volatility pair such as USDC-USDT. We refer to non-fragmented pools as \emph{single} (i.e., the unique pool for an asset pair).

We aggregate all interactions with Uniswap smart contracts into a panel across days and liquidity pools. To compute the end-of-day pool size, we account for all changes in token balances, across all price ranges. There are three possible interactions: A deposit or ``mint'' adds tokens to the pool, a withdrawal or ``burn'' removes tokens, whereas a trade or ``swap'' adds one token and removes the other. We track these changes across to obtain daily variation in the quantity of tokens on each pool. We obtain dollar values for the end-of-day liquidity pool sizes, intraday trade volumes, and liquidity events from the Uniswap V3 Subgraph. To determine a token's price in dollars, the Subgraph  searches for the most liquid path on Uniswap pools to establish the token's price in Ether and subsequently converts the Ether price to US dollars.

Table \ref{tab:sumstat} reports summary statistics across pools with different fee levels. High-fee pools attract on average 58\% of total liquidity supply, significantly more than their low-fee counterparts (\$46.50 million and \$33.78 million, respectively), but only capture 20.74\% percent of the trading volume (computed as $\nicefrac{8,071.24}{(8,071.24+30,848.79)}$ from the first column of Table \ref{tab:sumstat}). Consistent with our theoretical predictions, low-fee pools attract five times as many trades as high-fee competitors (610 versus 110 average trade count per day).  At the same time, the average trade on a high-fee pool is twice as large (\$14,490) than on a low-fee pool (\$6,340). 

The distribution of mint sizes is heavily skewed to the right, with 6.6\% of deposits exceeding \$1 million. There are large differences across pools -- the median $\LP$ deposit on the low-fee pool is \$15,680, twice as much as the median deposit on the high-fee pool (\$7,430). At the same time, the number of liquidity providers on high-fee pools is 51\% higher than on low-fee pools (10.08 unique addresses per day on high-fee pools versus only 6.68 unique address on high-fee pools).

\begin{table} \centering 
\caption{Descriptive statistics} 
  \label{tab:sumstat} 

\begin{minipage}[t]{1\columnwidth}%
\footnotesize
This table reports descriptive statistics for variables used in the empirical analysis. \emph{Pool size} is defined as the total value locked in the pool's smart contract at the end of each day. We compute the balance on day $t$ as follows: we take the balance at day $t-1$ and add (subtract) liquidity deposits (withdrawals) on day $t$, as well as accounting for token balance changes due to trades. The liquidity balance on the first day of the pool is taken to be zero. End of day balances are finally converted to US dollars. \emph{Daily volume} is computed as the sum of US dollar volume for all trades in a given pool and day. \emph{Liquidity share} (\emph{Volume share}) is computed as the ratio between a pool size (trading volume) for a given fee level and the aggregate size of all pools (trading volumes) for the same pair in a given day. \emph{Trade size} and \emph{Mint size} are the median trade and liquidity deposit size on a given pool and day, denominated in US dollars. \emph{Trade count} represents the number of trades in a given pool and day. \emph{\textbf{LP} wallets} counts the unique number of wallet addresses interacting with a given pool in a day.  The \emph{liquidity yield} is computed as the ratio between the daily trading volume and end-of-day TVL, multiplied by the fee tier. The \emph{price range} for every mint is computed as the difference between the top and bottom of the range, normalized by the range midpoint -- a measure that naturally lies between zero and two. \emph{Loss-versus-rebalancing} is computed as the permanent price impact of swaps with a one-hour horizon. The \emph{impermanent loss} is computed as in \citet{Heimbach2023} for a position in the range of 95\% to 105\% of the current pool price, with a forward-looking horizon of one hour. Finally, \emph{mint-to-burn} and \emph{burn-to-mint} times are defined as the time between a mint (burn) and a subsequent burn (mint) by the same address in the same pool, measured in hours. \emph{Mint-to-burn} and \emph{burn-to-mint} are recorded on the day of the final interaction with the pool. 
\end{minipage}

\vspace{0.05in}
\resizebox{0.91\textwidth}{!}{

\small
\begin{tabular}{@{\extracolsep{5pt}}llrrrrrr} 
\\[-1.8ex]\hline 
\hline \\[-1.8ex] 
Statistic & Pool fee & \multicolumn{1}{c}{Mean} & \multicolumn{1}{c}{Median} & \multicolumn{1}{c}{St. Dev.} & \multicolumn{1}{c}{Pctl(25)} & \multicolumn{1}{c}{Pctl(75)} & \multicolumn{1}{c}{N} \\ 
\hline \\[-1.8ex] 

Pool size (\$M) & Low & 33.78 & 2.05 & 96.91 & 0.30 & 14.12 & 20,151 \\   
& High & 46.50 & 3.85 & 95.73 & 1.43 & 27.51 & 20,151 \\ 
& Single & 3.89 & 0.84 & 13.56 & 0.26 & 2.62 & 130,767 \\ 

Liquidity share (\%) & Low & 39.52 & 35.52 & 32.53 & 7.37 & 72.16 & 20,151 \\ 
 & High & 60.48 & 64.48 & 32.53 & 27.84 & 92.63 & 20,151 \\

Daily volume (\$000) & Low & 30,848.79 & 619.77 & 118,908.80 & 6.18 & 5,697.30 & 20,151 \\  
& High & 8,071.24 & 114.96 & 36,777.38 & 7.83 & 1,882.12 & 20,151 \\  
& Single  & 915.73 & 36.07 & 6,059.78 & 1.93 & 277.00 & 130,767 \\    

Volume share & Low & 66.51 & 88.38 & 38.50 & 29.43 & 98.48 & 18,001 \\  
& High &  42.20 & 23.83 & 41.18 & 3.19 & 95.03 & 18,058 \\ 

Trade size (\$000) & Low & 6.34 & 2.20 & 13.36 & 0.61 & 6.03 & 18,001 \\ 
& High & 14.49 & 2.76 & 33.19 & 0.82 & 10.48 & 18,060 \\    
& Single    & 4.12 & 1.32 & 11.03 & 0.45 & 3.79 & 113,362 \\ 

Mint size (\$000) & Low & 820.84 & 15.68 & 13,114.83 & 3.78 & 58.98 & 10,640 \\  
& High & 1,001.10 & 7.43 & 13,807.10 & 1.55 & 30.52 & 10,370 \\ 
& Single & 96.97 & 6.93 & 622.12 & 1.42 & 30.39 & 45,300 \\ 

Trade count & Low & 610.61 & 95 & 1,518.52 & 12 & 414 & 20,151 \\   
& High & 110.59 & 26 & 490.29 & 8 & 89 & 20,151 \\  
& Single & 63.94 & 19 & 194.03 & 4 & 55 & 130,767 \\  

$\LP$ wallets & Low & 6.68 & 1 & 16.01 & 0 & 6 & 20,151 \\    
& High & 10.08 & 1 & 37.79 & 0 & 5 & 20,151 \\    
& Single  & 1.57 & 1.17 & 1.19 & 1.00 & 1.85 & 55,580 \\   

Liquidity yield (bps) & Low & 11.72 & 2.58 & 56.31 & 0.16 & 9.08 & 20,122 \\ 
& High & 9.69 & 1.65 & 51.44 & 0.15 & 6.40 & 20,130 \\
& Single & 17.90 & 1.94 & 90.18 & 0.18 & 8.58 & 130,433 \\

Price range & Low &  0.39 & 0.30 & 0.37 & 0.13 & 0.56 & 11,866 \\ 
& High & 0.61 & 0.54 & 0.44 & 0.32 & 0.84 & 12,195 \\    
& Single  & 0.68 & 0.58 & 0.52 & 0.27 & 1.02 & 55,580 \\ 

Loss-versus-rebalancing (bps) & Low & 14.24 & 1.32 & 35.20 & 0.02 & 9.22 & 20,151 \\ 
& High & 7.85 & 0.84 & 23.15 & 0.03 & 4.87 & 20,151 \\

Impermanent loss (bps) & Low  & 8.46 & 1.84 & 27.88 & 0.06 & 7.23 & 20,118 \\  
& High & 7.37 & 1.33 & 27.21 & 0.05 & 5.93 & 20,132 \\  
& Single  & 17.20 & 2.44 & 71.34 & 0.17 & 11.37 & 130,340 \\  

Mint-to-burn (hrs) & Low & 450.40 & 59.82 & 1,341.67 & 19.70 & 243.83 & 10,186 \\  
& High & 952.14 & 165.61 & 2,076.42 & 39.66 & 711.67 & 9,979 \\   
& Single & 760.26 & 126.64 & 1,778.62 & 27.01 & 563.50 & 39,735 \\   

Burn-to-mint (hrs) & Low & 105.29 & 0.20 & 521.62 & 0.08 & 5.63 & 8,279 \\ 
& High & 224.26 & 0.32 & 941.31 & 0.10 & 27.78 & 7,289 \\ 
& Single & 177.74 & 0.23 & 803.40 & 0.07 & 20.64 & 27,477 \\  

\hline \\[-1.8ex] 

\end{tabular} 
}
\end{table} 

One concern with measuring average mint size is just-in-time liquidity provision (JIT). As discussed for example in \citet{capponi2024paradox}, JIT liquidity providers submit very large and short-lived deposits to the pool to dilute competitors on an incoming large trade; they immediately withdraw the balance in the same block after executing the trade. In our sample, JIT liquidity provision is not economically significant, accounting for less than 1\% of aggregate trading volume. However, it has the potential to skew mint sizes to the right, particularly in low-fee pools, without providing liquidity to the market at large. We address this issue by (i) filtering out JIT mints using the algorithm in Appendix \ref{sec:app-jit} and (ii) taking the median liquidity mint size at day-pool level rather than the mean. 

Further, we follow \citet{augustin2022reaching} to compute the daily liquidity fee yield as the product between pool's fee tier and the ratio between trading volume and the lagged total value locked (TVL). That is,
\begin{equation}\label{eq:liq_yield}
    \text{Liquidity yield}=\text{liquidity fee}_i \times \frac{\text{Volume}_{i,t}}{\text{TVL}_{i,t-1}},
\end{equation}
for pool $i$ and day $t$. The average daily yield is slightly higher on low-fee pools, at 11.72 basis points, compared to 9.69 basis points on high-fee pools.

A salient observation in Table \ref{tab:sumstat} is that non-fragmented pairs (``single'' pools) are significantly smaller -- on average less than 10\% of the pool size and trading volume of fragmented pairs. Average trade and mint sizes are correspondingly lower as well. The evidence suggests that pairs for which there is significant trading interest, and therefore potentially a broader cross-section of potential liquidity providers, are more likely to become fragmented.

\begin{figure}
\caption{\label{fig:stat_liq} Liquidity supply on decentralized exchanges}
\begin{minipage}[t]{1\columnwidth}%
\footnotesize
This figure plots the empirical distributions of variables in the pool-day panel, across low and high fee pools (for fragmented pairs) as well as single pools in pairs that are not fragmented. In each box plot, the median is marked as a vertical line; the box extends to the quartiles of the data set, whereas the whiskers extend to an additional 1.5 times the inter-quartile range.
\end{minipage}

\vspace{0.05in}

\begin{centering}

\subfloat[Pool size and trading volume]{
\includegraphics[width=\textwidth]{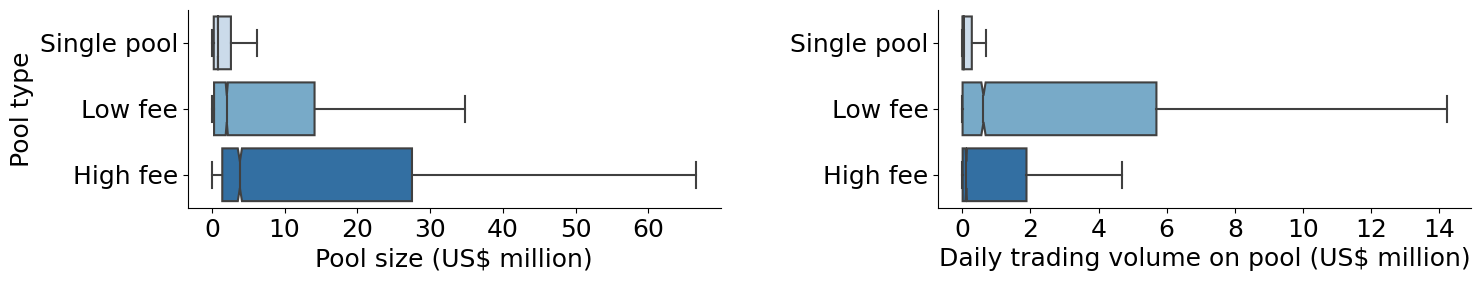}
}

\subfloat[Average liquidity mint and trade size]{
\includegraphics[width=\textwidth]{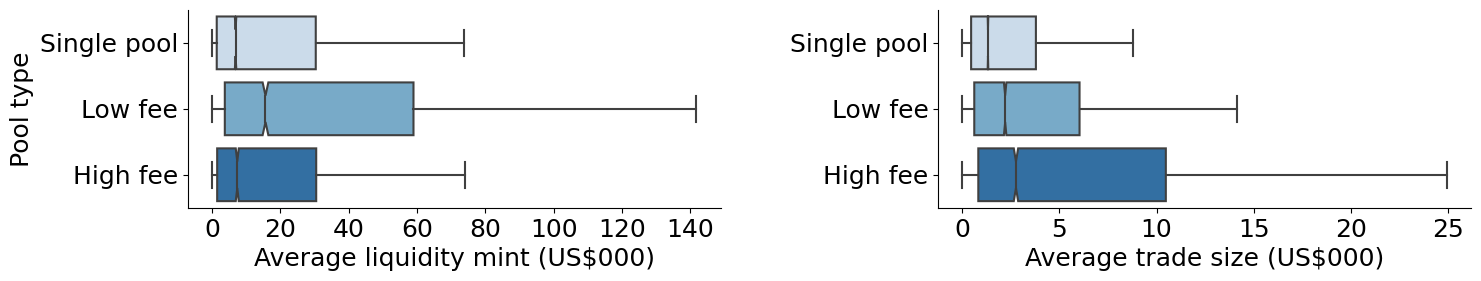}
}

\subfloat[Number of liquidity providers and trades]{
\includegraphics[width=\textwidth]{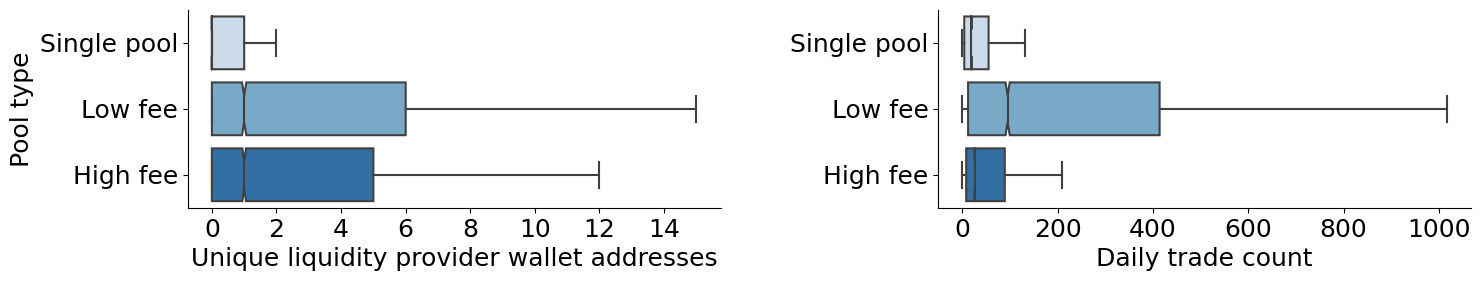}
}
\par\end{centering}

\end{figure}

Figure \ref{fig:stat_liq} plots the distributions of our empirical measures across low- and high-fee liquidity pools. It suggest a sharp segmentation of liquidity supply and trading across pools. High-fee pools attract smaller liquidity providers by mint size, and end up with a larger \emph{aggregate} size than their low-fee counterparts. Trading volume is similarly segmented: most small value trades are executed on the cheaper low-fee pools, making up the majority of daily volume for a given pair. High-value trades, of which there are fewer, are more likely to (also) execute on high-fee pools.

\begin{figure}[H]
\caption{\label{fig:liq_cycles} Liquidity cycles on high- and low-fee pools}
\begin{minipage}[t]{1\columnwidth}%
\footnotesize
The top panel plots the distribution of liquidity cycle times from mint to subsequent burn (left) and from burn to subsequent mint (right) for the same $\LP$ wallet address in the same pool. In each box plot, the median is marked as a vertical line; the box extends to the quartiles of the data set, whereas the whiskers extend to an additional 1.5 times the inter-quartile range. The bottom panel plots the probability that the $\LP$ position is out of range and therefore does not earn fees. A position is considered to be ``out of range'' when the minimum and maximum prices at which the $\LP$ is willing to provide liquidity do no straddle the current price on the pool. We plot the probability separately for low- and high- fees, as well as conditional on whether the event is a burn (liquidity withdrawal) or mint (liquidity deposit).
\end{minipage}

\vspace{0.05in}

\begin{centering}

\includegraphics[width=\textwidth]{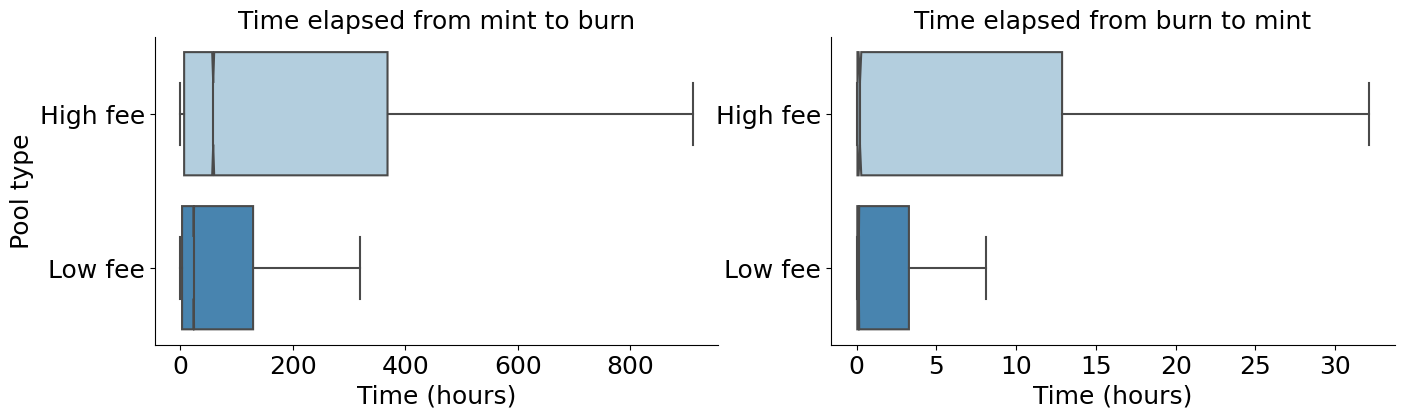}
\includegraphics[width=\textwidth]{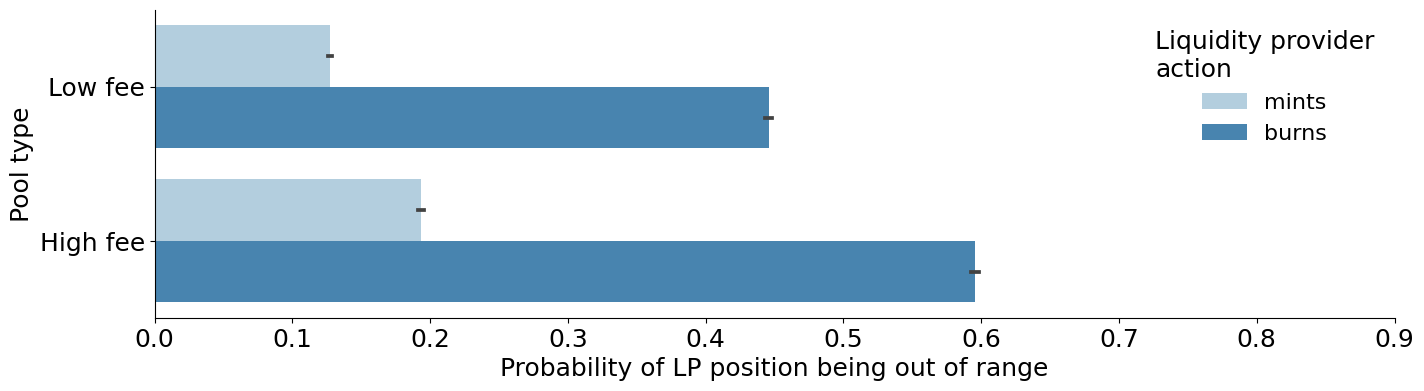}

\par\end{centering}

\end{figure}

Our theoretical framework in Section \ref{sec:model} implies that liquidity suppliers manage their positions more actively in the low- than the high-fee pool. Figure \ref{fig:liq_cycles} provides suggestive evidence for liquidity cycles of different lengths in the cross-section of pools. Liquidity on decentralized exchanges is significantly more passive than on traditional equity markets. That is, liquidity providers do not often manage their positions at high frequencies. The median time from a mint (deposit) to a subsequent burn (withdrawal) from the same wallet on the same pool ranges from 59.82 hours, or 2.49 days, on low-fee pools to 165.61 hours, or 6.9 days on high-fee pools. 

When do $\LP$s re-balance their positions? In 53\% of cases, liquidity providers only withdraw tokens from the pool when their position exits the price range that allows them to collect fees. Concretely, $\LP$s specified price range for liquidity provision does not straddle the most recent reference price of the pool. The scenario mirrors a limit order market where a liquidity provider's outstanding limit orders are deep in the book, such that she doesn't stand to earn the spread on the marginal incoming trade. In this case, a rational market maker might want to cancel their outstanding order and place a new one at the top of the book. This is exactly the pattern we observe on Uniswap: the subsequent mint following a burn straddles the new price 77\% of the time -- $\LP$s reposition
their liquidity around the current prices to keep earning fees on incoming trades. Moreover, re-balancing is swift -- the median time between a burn and a subsequent mint is just 12 minutes (0.20 hours).

The empirical pattern in Figure \ref{fig:liq_cycles} echoes the re-balancing cycles as described in Section \ref{sec:model}. Liquidity providers deposit tokens in Uniswap pools to trade against uninformed order flow. They only re-balance when their position becomes out-of-range and no longer earns fees. Once this happens, $\LP$s quickly adjust their position in a matter of minutes -- by removing stale liquidity and adding a new position around the current price. The re-balancing cycle tends to be longer on high-fee pools, where arbitrageurs only move the price outside the range if the asset value innovation is large enough.

We note that $\LP$s do not seem to ``race'' to update liquidity upon information arrival as in \citet{Budish2015TheResponse}. First, they very rarely manage their position intraday. Second, $\LP$s on Uniswap typically do not remove in-range liquidity that stands to trade first against incoming order flow and therefore bears the highest adverse selection risk. Our results are consistent with \citet{CapponiJia2021} who theoretically argue that $\LP$s have low incentives to compete with arbitrageurs on news arrival, as well as with \citet{CapponiJiaYu2022} who find no evidence of traders racing to trade on information on Uniswap v2. In our model, $\LP$s tend to re-balance their position \emph{after} an arbitrageur executed their trade.

\begin{figure}[H]
\caption{\label{fig:routing} Optimal order routing on Uniswap v3 pools}
\begin{minipage}[t]{1\columnwidth}%
\footnotesize
This figure displays the optimal order split for purchasing ETH using USDC across various trade sizes, on USDC-WETH Uniswap v3 pools with liquidity fees of 5 and 30 basis points. Order execution is optimized to minimize trading costs, encompassing both price impact and liquidity fees. We use liquidity distribution snapshots data from Kaiko, and focus on the last Ethereum block of each day from May 4, 2021, to July 15, 2023.
\end{minipage}

\vspace{0.05in}

\begin{centering}
\includegraphics[width=\textwidth]{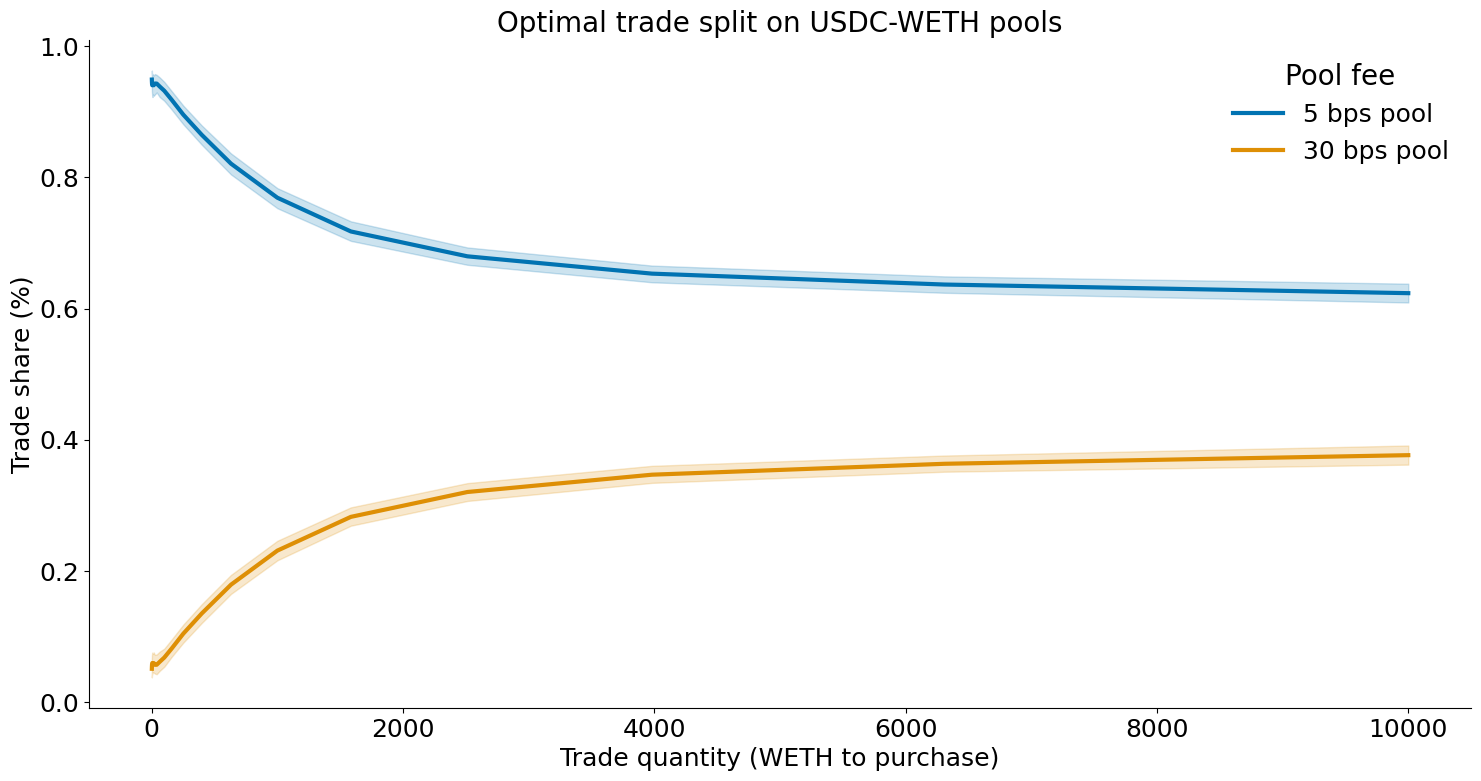}
\par\end{centering}

\end{figure}

Next, we examine the behavior of liquidity takers (\(\LT\)). According to our model, small orders are typically routed to low-fee pools, while larger orders are split between both low- and high-fee pools. Figure \ref{fig:routing} provides empirical evidence supporting this claim. We use liquidity snapshot data from Kaiko on USDC-WETH pools to simulate the optimal routing strategy for trades of various sizes for the last block of each day in our sample. This simulation considers both the price impact of trades and the associated liquidity fees. In line with our model, we find that trades smaller than 150 ETH (approximately \$450,000) optimally route over 90\% of their size to the low-fee pool. Conversely, larger trades distribute their volume more evenly, with up to 40\% being executed in high-fee pools.

\paragraph{Measuring gas prices.} Each interaction with smart contracts on the Ethereum blockchain requires computational resources, measured in units of ``gas.'' Upon submitting a mint or burn transaction to the decentralized exchange, each liquidity provider specifies their willingness to pay per unit of gas, that is they bid a  ``gas price.'' Traders are likely to bid higher prices for more complex transactions or if they require a faster execution. To generate a conservative daily benchmark for the gas price, we compute the average of the lowest 1000 user gas bids for mint and burn interactions on day $t$, across all liquidity pools in the benchmark sample. 

\begin{figure}
\caption{\label{fig:gascosts} Gas costs for Uniswap v3 mint/burn transactions}
\begin{minipage}[t]{1\columnwidth}%
\footnotesize
The figure illustrates the daily average gas cost on mint/burn transactions in Uniswap v3 pools. The gas cost is computed as the average of the lowest 1000 user gas bids for mint and burn interactions on each day, across all liquidity pools in the benchmark sample.
\end{minipage}

\vspace{0.05in}

\begin{centering}

\includegraphics[width=\textwidth]{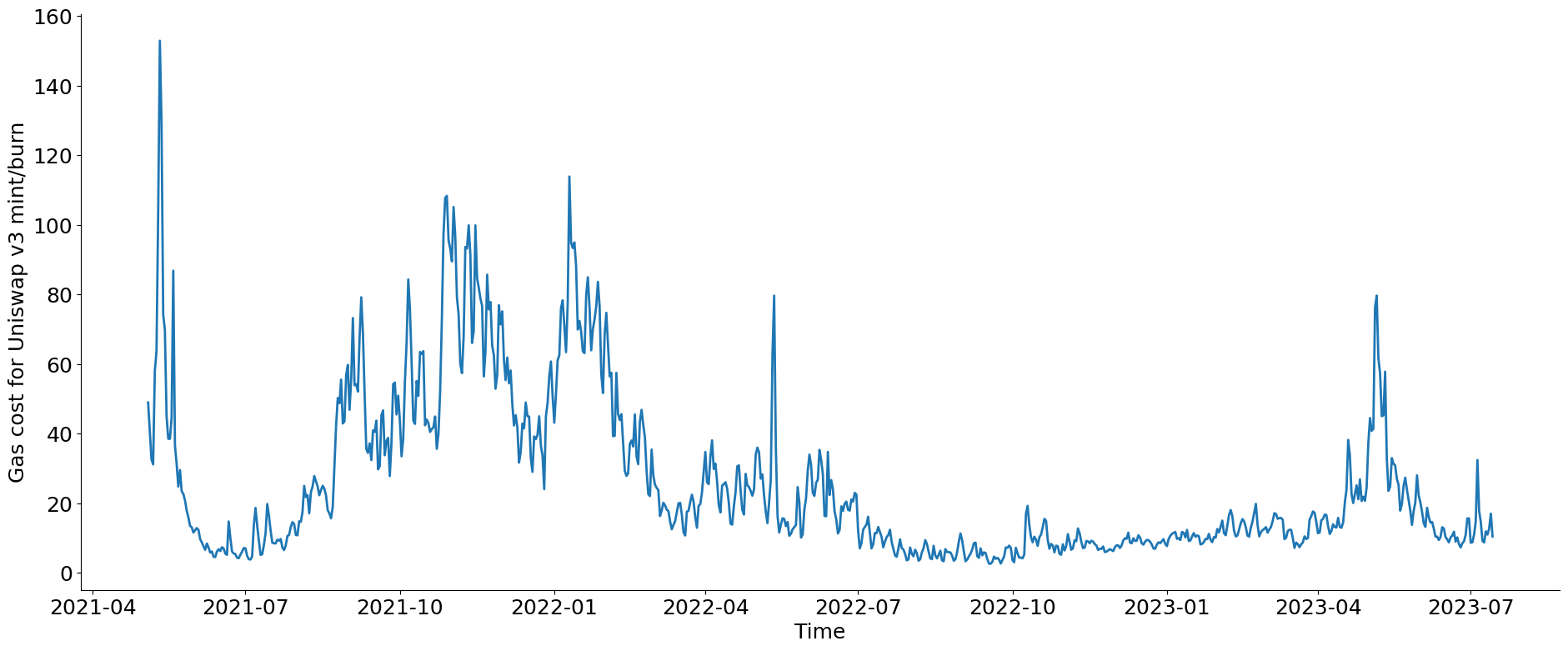}

\par\end{centering}

\end{figure}

Figure \ref{fig:gascosts} showcases the significant fluctuation in gas costs for Uniswap liquidity transactions over time. 
Gas costs denominated in USD are influenced by two primary factors: network congestion, which leads to variations in gas prices measured in Ether, and the fluctuation of Ether's value relative to the US dollar. On a monthly average, gas costs peaked at above US\$100 in November 2021 and have since plummeted to around US\$6 from the second half of 2022, albeit with occasional spikes.

\section{Empirical results}
\subsection{Liquidity supply on high- and low-fee pools}

To formally test the model predictions and quantify the differences in liquidity supply across fragmented pools, we build a panel data set for the 32 fragmented pairs in our sample where the unit of observation is pool-day. We estimate linear regressions of liquidity and volume measures on liquidity fees and gas costs:
\begin{equation}
    y_{ijt}=\alpha + \beta_0 d_\emph{low-fee, ij} + \beta_1 \text{GasPrice}_{jt} + \beta_2 \text{GasPrice}_{jt} \times d_\emph{low-fee, ij} + \sum \beta_k \emph{Controls}_{ijt} + \theta_j + \delta_w + \varepsilon_{ijt},
\end{equation}
where $y$ is a variable of interest, $i$ indexes liquidity pools, $j$ runs over asset pairs, and $t$ and $w$ indicates days and weeks, respectively. The dummy $d_\emph{low-fee, ij}$ takes the value one for the pool with the lowest fee in pair $j$ and zero else. 

Further, our set of controls includes pair and week fixed effects, the log aggregate trading volume and log liquidity supply (i.e., total value locked) for day $t$ across all pools $i$. Volume and liquidity are measured in US dollars. We also control for daily return volatility, computed as the range between the daily high and low prices for a given pair $j$ \citep[following][]{Alizadeh2002}: 
\begin{equation}
    \text{Volatility}_{jt}=\frac{1}{2\sqrt{\log 2}} \log\left(\frac{\text{High}_{jt}}{\text{Low}_{jt}}\right).
\end{equation}
To measure volatility for fragmented pairs that actively trade in multiple pools, we select the pool with the highest trading volume for a given day.

Consistent with Figure \ref{fig:stat_liq}, we show in Table \ref{tab:markeshare} that most of the capital deployed to provide liquidity for a given pair is locked in high-fee pools. At the same time, low-fee pools attract much larger trading volume. Models (1) and (5) show that the average low-fee pool attracts 39.5\% of liquidity supply for the average pair (that is, equal to $\nicefrac{(100-20.92)}{2}$) while it executes 62\% (i.e., $\nicefrac{(100+24.62)}{2}$) of the total trading volume. At a first glance, it would seem that a majority of capital on decentralized exchanges is inefficiently deployed in pools with low execution probability. We will show that, in line with our model, the difference is driven by heterogeneous rebalancing costs across pools, leading to the formation of $\LP$ clienteles. 

The regression results in Table \ref{tab:markeshare} support Prediction \ref{pred:comp_stat_Gamma}, stating that market share differences between pools are linked to variation in fixed transaction costs on the blockchain. A one-standard deviation increase in gas prices leads to a 4.63 percentage point increase in the high-fee liquidity share. The results suggests that blockchain transaction costs have an economically meaningful and statistically significant impact on liquidity fragmentation. In line with the theoretical model in Section \ref{sec:model}, a jump in gas prices leads to a reshuffling of liquidity supply from low- to high-fee pools. 

Evidence suggests that a higher gas price leads to a 6.52\% lower volume share for the low-fee pool. This outcome is natural, as the incoming order flow is optimally routed to the high-fee pool, following the liquidity providers.


\begin{table}
  \caption{Liquidity pool market shares and gas prices}   \label{tab:markeshare}
\begin{minipage}[t]{1\columnwidth}%
\footnotesize
			This table reports the coefficients of the following regression:
	\begin{align*}
    \text{MarketShare}_{ijt}=\alpha + \beta_0 d_\emph{low-fee, ij} + \beta_1 \text{GasPrice}_{jt} + \beta_2 \text{GasPrice}_{jt} \times d_\emph{low-fee, ij} + \sum \beta_k \emph{Controls}_{ijt} + \theta_j + \varepsilon_{ijt}
    \end{align*}
 	where the dependent variable is the liquidity or trading volume market share for pool $i$ in asset pair $j$ on day $t$. $d_\emph{low-fee, ij}$ is a dummy that takes the value one for the pool with the lowest fee in pair $j$ and zero else. $\emph{GasPrice}_{jt}$ is the average of the lowest 100 bids on liquidity provision events across all pairs on day $t$, standardized to have a zero mean and unit variance. \emph{Volume} is the natural logarithm of the sum of all swap amounts on day $t$, expressed in thousands of US dollars. \emph{Total value locked} is the natural logarithm of the total value locked on Uniswap v3 pools on day $t$, expressed in millions of dollars.\emph{Volatility} is computed as the daily range between high and low prices on the most active pool for a given pair.All regressions include pair and week fixed-effects. Robust standard errors in parenthesis are clustered by week and ***, **, and * denote the statistical significance at the 1, 5, and 10\% level, respectively.  The sample period is from May 4, 2021 to July 15, 2023. 
\end{minipage}
\small
\begin{center}
\resizebox{\textwidth}{!}{
\begin{tabular}{lcccc@{\hskip 0.3in}cccc}
\toprule
 & \multicolumn{4}{c}{Liquidity market share (\%)} & \multicolumn{4}{c}{Volume market share (\%)} \\ 
 & (1) & (2) & (3) & (4) & (5) & (6) & (7) & (8) \\
 \cmidrule{1-9}

$ d_\text{low-fee}$ & -20.92*** & -20.92*** & -20.92*** & -20.94*** & 24.62*** & 24.63*** & 24.62*** & 24.71*** \\
 & (-27.42) & (-27.41) & (-27.42) & (-23.95) & (20.55) & (20.56) & (20.55) & (18.54) \\
Gas price $\times$ $ d_\text{low-fee}$ & -4.63*** & -4.62*** & -4.63*** &  & -6.52*** & -6.52*** & -6.52*** &  \\
 & (-7.32) & (-7.32) & (-7.32) &  & (-5.92) & (-5.92) & (-5.92) &  \\
Gas price & 2.31*** & 2.31*** & 2.31*** &  & 3.63*** & 3.61*** & 3.61*** &  \\
 & (7.32) & (7.32) & (7.32) &  & (7.33) & (7.30) & (7.26) &  \\
Volume & 0.00 & 0.00 & 0.00 & 0.00 & -0.19** & -0.20** & -0.19** & -0.12 \\
 & (0.65) & (1.33) & (0.65) & (0.66) & (-2.54) & (-2.61) & (-2.50) & (-1.56) \\
Total value locked & -0.00 & -0.00 &  & -0.00 & 0.58 & 0.58 &  & 0.44 \\
 & (-0.58) & (-0.06) &  & (-0.64) & (1.44) & (1.44) &  & (1.10) \\
Volatility & -0.29 &  & -0.29 & -0.28 & -1.15*** &  & -1.15*** & -1.13** \\
 & (-0.90) &  & (-0.90) & (-0.82) & (-2.74) &  & (-2.74) & (-2.56) \\
Constant & 60.45*** & 60.46*** & 60.45*** & 60.46*** & 41.96*** & 41.99*** & 41.96*** & 41.96*** \\
 & (158.00) & (158.46) & (158.00) & (137.54) & (69.99) & (70.22) & (70.02) & (62.81) \\
Pair FE & Yes & Yes & Yes & Yes & Yes & Yes & Yes & Yes \\
Week FE & Yes & Yes & Yes & Yes & Yes & Yes & Yes & Yes \\
Observations & 40,288 & 40,288 & 40,288 & 40,288 & 36,059 & 36,059 & 36,059 & 36,059 \\
 R-squared & 0.10 & 0.10 & 0.10 & 0.09 & 0.13 & 0.13 & 0.13 & 0.12 \\ \hline
\bottomrule
\multicolumn{9}{l}{Robust t-statistics in parentheses. Standard errors are clustered at week level.  *** p$<$0.01, ** p$<$0.05, * p$<$0.1} \\
\end{tabular}
}
\end{center}
\end{table} 

What drives the market share gap across fragmented pools? In Table \ref{tab:ordersize} we document stark differences between the characteristics of individual orders supplying or demanding liquidity on pools with low and high fees. On the liquidity supply side, model (1) in Table \ref{tab:ordersize} shows that the average liquidity mint is 107.5\% larger on low-fee pools, which supports Prediction \ref{pred:clienteles} of the model.\footnote{Since all dependent variables are measured in natural logs, the marginal impact of a dummy coefficient $\beta$ is computed $\left(e^\beta-1\right) \times 100$ percent.} At the same time, there are 3.40 fewer unique wallets (Model 5) providing liquidity on the low-fee pool -- that is, a 34\% relative difference between high- and low-fee pools.

\begin{table}
\caption{Fragmentation and order flow characteristics}  \label{tab:ordersize}
\begin{minipage}[t]{1\columnwidth}%
\footnotesize
			This table reports the coefficients of the following regression:
	\begin{align*}
    y_{ijt}=\alpha + \beta_0 d_\emph{low-fee, ij} + \beta_1 \text{GasPrice}_{jt} d_\emph{low-fee, ij} + \beta_2 \text{GasPrice}_{jt} \times d_\emph{high-fee, ij} + \sum \beta_k \emph{Controls}_{ijt} + \theta_j + \varepsilon_{ijt}
    \end{align*}
	where the dependent variable $y_{ijt}$ can be (i) the log median mint size, (ii) the log median trade size, (iii) the log trading volume, (iv) the log trade count $\log(1+\# trades)$, (v) count of unique $\LP$ wallets interacting with a pool in a given day, (vi) the liquidity yield in bps for pool $i$ in asset $j$ on day $t$, computed as in equation \eqref{eq:liq_yield}, and (vii) the average liquidity mint price range for pool $i$ in asset $j$ on day $t$. Price range is computed as the difference between the top and bottom of the range, normalized by the range midpoint -- a measure that naturally lies between zero and two. $d_\emph{low-fee, ij}$ is a dummy that takes the value one for the pool with the lowest fee in pair $j$ and zero else. $d_\emph{high-fee, ij}$ is defined as $1-d_\emph{low-fee, ij}$. $\emph{GasPrice}_{jt}$ is the average of the lowest 100 bids on liquidity provision events across all pairs on day $t$, standardized to have a zero mean and unit variance. \emph{Volume} is the natural logarithm of the sum of all swap amounts on day $t$, expressed in thousands of US dollars. \emph{Total value locked} is the natural logarithm of the total value locked on Uniswap v3 pools on day $t$, expressed in millions of dollars. \emph{Volatility} is computed as the daily range between high and low prices on the most active pool for a given pair. All regressions include pair and week fixed-effects. Robust standard errors in parenthesis are clustered by week and ***, **, and * denote the statistical significance at the 1, 5, and 10\% level, respectively.  The sample period is from May 4, 2021 to July 15, 2023. 
\end{minipage}
\begin{center}
\resizebox{1\textwidth}{!}{  
\begin{tabular}{lccccccc}
\toprule
 & \multicolumn{1}{c}{Mint size} &  \multicolumn{1}{c}{Trade size} & \multicolumn{1}{c}{Volume} & \multicolumn{1}{c}{\# Trades}  & \multicolumn{1}{c}{\# Wallets} & \multicolumn{1}{c}{Liquidity yield} & \multicolumn{1}{c}{Price range} \\
 & (1) & (2) & (3) & (4) & (5) & (6) & (7) \\
\cmidrule{1-8}
$ d_\text{low-fee}$ & 0.73*** & -0.30*** & 0.89*** & 1.02*** & -3.40*** & 2.03*** & -0.18*** \\
 & (12.27) & (-10.05) & (14.23) & (32.95) & (-5.00) & (3.60) & (-41.84) \\
Gas price $\times$ $ d_\text{low-fee}$ & 0.37*** & 0.08*** & -0.03 & -0.22*** & -3.00*** & 3.57** & -0.00 \\
 & (4.96) & (3.75) & (-0.95) & (-7.29) & (-3.43) & (2.30) & (-0.47) \\
Gas price $\times$ $ d_\text{high-fee}$ & 0.58*** & 0.17*** & 0.24*** & 0.07** & -2.89*** & 5.57*** & -0.03*** \\
 & (7.52) & (8.81) & (5.95) & (2.46) & (-3.15) & (2.83) & (-4.65) \\
Volume & 0.37*** & 0.16*** & 0.43*** & 0.20*** & 1.22*** & 1.01 & -0.01** \\
 & (8.68) & (21.38) & (15.27) & (13.85) & (6.56) & (0.81) & (-2.56) \\
Total value locked & -0.16 & 0.11*** & 0.23** & -0.01 & -1.86 & -13.42 & -0.02 \\
 & (-1.30) & (3.54) & (1.99) & (-0.18) & (-0.99) & (-1.09) & (-0.99) \\
Volatility & -0.04 & -0.01 & -0.07 & 0.01 & -0.09 & 1.18** & 0.02*** \\
 & (-1.11) & (-1.34) & (-1.38) & (0.88) & (-1.03) & (2.21) & (3.98) \\
Constant & 1.88*** & 1.64*** & 5.27*** & 3.26*** & 10.12*** & 10.01*** & 0.59*** \\
 & (58.27) & (111.47) & (168.58) & (209.84) & (28.65) & (26.04) & (184.91) \\
 &  &  &  &  &  &  \\
Pair FE & Yes & Yes & Yes & Yes & Yes & Yes & Yes  \\
Week FE & Yes & Yes & Yes & Yes & Yes & Yes  & Yes \\
Observations & 21,000 & 36,059 & 36,059 & 40,288 & 40,288 & 40,252 & 24,058 \\
 R-squared & 0.26 & 0.53 & 0.55 & 0.52 & 0.37 & 0.09 & 0.42 \\ \hline
\bottomrule
\multicolumn{7}{l}{Robust t-statistics in parentheses. Standard errors are clustered at week level.} \\
\multicolumn{7}{l}{*** p$<$0.01, ** p$<$0.05, * p$<$0.1} \\
\end{tabular}
}
\end{center}
\end{table}

On the liquidity demand side, trades on the low-fee pool are 25.91\% smaller (Model 2), consistent with Prediction \ref{pred:trade_size_volume}. However, the low-fee pool executes almost three times the number of trades (i.e., trade count is 177\% higher from Model 4) and has 143\% higher volume than the high-fee pool (Model 3). 

Next, in line with Prediction \ref{pred:ly}, low-fee pools generate a higher liquidity yield. On average, liquidity providers on low-fee pools earn 2.03 basis points higher revenue than their counterparts on high-fee pools (Model 6), indicating significant positive returns resulting from economies of scale.

Our findings (Model 7) indicate that liquidity providers on low-fee pools select price ranges that are 30\% (=0.18/0.59) narrower when minting liquidity compared to those on high-fee pools. This pattern aligns with the capability of large LPs to adjust their liquidity positions frequently, enabling more efficient capital concentration. Similarly, \citet{caparros2023blockchain} report a higher concentration of liquidity in pools on alternative blockchains like Polygon, known for lower transaction costs than Ethereum.

The results point to an asymmetric match between liquidity supply and demand across pools. On low-fee pools, a few $\LP$s provide large chunks of liquidity for the vast majority of incoming small trades. Conversely, on high-fee pools there is a sizeable mass of small liquidity providers that mostly trade against a few large incoming trades. 

How does variation in fixed transaction costs impact the gap between individual order size across pools? We find that increasing the gas price by one standard deviation leads to higher liquidity deposits on both the low- and the high-fee pools (14.2\% and 30.1\% higher, respectively).\footnote{The relative effects are computed as $\nicefrac{0.37}{(1.88+0.73)}=13.8\%$ for low pools and $\nicefrac{0.58}{1.88}=30.85\%$ for high-fee pools, respectively.} The result supports Prediction \ref{pred:clienteles_cs} of the model. Our theoretical framework implies that a larger gas price leads to some (marginal) $\LP$s switching from the low- to the high-fee pool. The switching $\LP$s have low capital endowments relative to their low-fee pool peers, but higher than $\LP$s on the high-fee pool. Therefore, the gas-driven reshuffle of liquidity leads to a higher average endowment on both high- and low-fee pools. Consistent with the model, a higher gas price leads to fewer active liquidity providers, particularly on low-fee pools. Specifically, a one-standard increase in gas costs leads to a significant decrease in the number of $\LP$ wallets interacting daily with low- and high-fee pools, respectively (Model 5). 

While a higher gas price is correlated with a shift in liquidity supply, it has a muted impact on liquidity demand on low-fee pools. A higher gas cost is associated with 7.6\% larger trades (Model 2), likely as traders aim to achieve better economies of scale. At the same time, the number of trades on the low-fee pool drops by 19.7\% (Model 4) -- since small traders might be driven out of the market. The net of gas prices effect on aggregate volume on the low-fee pool is small and not statistically significant (Model 3). The result matches our model assumption that the aggregate order flow on low-fee pool is not sensitive to gas prices.

On the high-fee pool, a higher gas price is also associated with a higher trade size, but also an increase in traded volume. As gas prices rise, liquidity providers switch from low- to high-fee pools. The outcome is greater depth and reduced price impact for liquidity demanders on high fee pools, which leads to higher trading volume.

\begin{table}[H]
\caption{Liquidity flows and gas costs on fragmented pools} \label{tab:flows}
\begin{minipage}[t]{1\columnwidth}%
\footnotesize
			This table reports the coefficients of the following regression:
	\begin{align*}
    y_{ijt}=\alpha + \beta_0 d_\emph{low-fee, ij} + \beta_1 \text{GasPrice}_{jt} d_\emph{low-fee, ij} + \beta_2 \text{GasPrice}_{jt} \times d_\emph{high-fee, ij} + \sum \beta_k \emph{Controls}_{ijt} + \theta_j + \varepsilon_{ijt}
    \end{align*}
	where the dependent variable $y_{ijt}$ can be (i) the aggregate dollar value of mints (in logs), or (vi) a dummy variable taking value one hundred if there is at least one mint on liquidity pool $i$ in asset $j$ on day $t$. $d_\emph{low-fee, ij}$ is a dummy that takes the value one for the pool with the lowest fee in pair $j$ and zero else. $d_\emph{high-fee, ij}$ is defined as $1-d_\emph{low-fee, ij}$. $\emph{GasPrice}_{jt}$ is the average of the lowest 100 bids on liquidity provision events across all pairs on day $t$, standardized to have a zero mean and unit variance. \emph{Volume} is the natural logarithm of the sum of all swap amounts on day $t$, expressed in thousands of US dollars. \emph{Total value locked} is the natural logarithm of the total value locked on Uniswap v3 pools on day $t$, expressed in millions of dollars.\emph{Volatility} is computed as the daily range between high and low prices on the most active pool for a given pair.All regressions include pair and week fixed-effects. Robust standard errors in parenthesis are clustered by week and ***, **, and * denote the statistical significance at the 1, 5, and 10\% level, respectively.  The sample period is from May 4, 2021 to July 15, 2023. 
\end{minipage}
\begin{center}
\begin{tabular}{lcccccc}
\toprule
 & \multicolumn{3}{c}{Daily mints (log US\$)} &  \multicolumn{3}{c}{$\text{Prob}\left(\text{at least one mint}\right)$} \\
 & (1) & (2) & (3) & (4) & (5) & (6) \\
\cmidrule{1-7}
$ d_\text{low-fee}$ & 0.43*** & 0.43*** & 0.43*** & 1.38* & 1.37* & 1.38* \\
 & (6.07) & (6.07) & (6.07) & (1.71) & (1.71) & (1.71) \\
Gas price $\times$ $ d_\text{low-fee}$ & -0.35*** & -0.35*** & -0.46*** & -6.02*** & -6.01*** & -4.58*** \\
 & (-8.50) & (-8.50) & (-7.14) & (-9.13) & (-9.13) & (-6.76) \\
Gas price $\times$ $ d_\text{high-fee}$ & 0.11** & 0.11** &  & -1.43** & -1.43** &  \\
 & (2.15) & (2.15) &  & (-2.57) & (-2.57) &  \\
Volume & 0.26*** & 0.26*** & 0.26*** & 0.96*** & 0.96*** & 0.96*** \\
 & (14.78) & (14.77) & (14.78) & (3.93) & (3.93) & (3.93) \\
Total value locked & -0.07 & -0.07 & -0.07 & 1.47 & 1.47 & 1.47 \\
 & (-0.78) & (-0.78) & (-0.78) & (1.01) & (1.00) & (1.01) \\
Volatility & -0.01 &  & -0.01 & 0.26 &  & 0.26 \\
 & (-0.68) &  & (-0.68) & (0.59) &  & (0.59) \\
Gas price &  &  & 0.11** &  &  & -1.43** \\
 &  &  & (2.15) &  &  & (-2.57) \\
Constant & 2.61*** & 2.61*** & 2.61*** & 51.44*** & 51.43*** & 51.44*** \\
 & (73.46) & (73.49) & (73.46) & (126.67) & (127.28) & (126.67) \\
 &  &  &  &  &  &  \\
Pair FE & Yes & Yes & Yes & Yes & Yes & Yes  \\
Week FE & Yes & Yes & Yes & Yes & Yes & Yes  \\
Observations & 40,288 & 40,288 & 40,288 & 40,288 & 40,288 & 40,288 \\
 R-squared & 0.47 & 0.47 & 0.47 & 0.28 & 0.28 & 0.28 \\ \hline
\bottomrule
\multicolumn{7}{l}{Robust t-statistics in parentheses. Standard errors are clustered at week level.} \\
\multicolumn{7}{l}{*** p$<$0.01, ** p$<$0.05, * p$<$0.1} \\
\end{tabular}
\end{center}
\end{table}

In Table \ref{tab:flows}, we shift the analysis from individual orders to aggregate daily liquidity flows to Uniswap pools. We find that higher gas prices lead to a decrease in liquidity inflows, but only on the low fee pools. A one standard deviation increase in gas prices leads to a 29.5\% drop in new liquidity deposits by volume (Model 1) and an 6.02\% drop in probability of having at least one mint (Model 4) on the low-fee pool. However, the slow-down in liquidity inflows is less evident in high fee pools. While an increase in gas prices reduce the probability of liquidity inflows by 1.43\%, it actually leads to a 11.6\% increase in the daily dollar inflow to the pool. Together with the result in Table \ref{tab:ordersize} that the size of individual mints increases with gas prices, our evidence is consistent with the model implication that higher fixed transaction costs change the composition of liquidity supply on the high-fee pool, with small $\LP$ being substituted by larger $\LP$s switching over from the low-fee pool.

\subsection{Re-balancing activity on high- and low-fee pools}
Next, we test Prediction \ref{pred:updates} on the duration of liquidity re-balancing cycles on fragmented pools. Since the descriptive statistics in Table \ref{tab:sumstat} suggest that $\LP$s manage their positions over multiple days, we cannot accurately measure liquidity cycles in a pool-day panel. Instead, we use intraday data on liquidity events (either mints or burns) to measure the duration between two consecutive opposite-sign interactions by the same Ethereum wallet with a liquidity pool: either a mint followed by a burn, or vice-versa. 

To ensure consistency with the model described in Section \ref{sec:model}, we conduct our analysis on the entire sample as well as on a sub-sample focused solely on re-balancing events where the liquidity position falls out of range (i.e., the price range set by the $\LP$ does not straddle the current price and therefore the $\LP$ does not earn fees). We further introduce wallet fixed effects to soak up variation in reaction times across traders, and winsorize the liquidity cycle duration at the 1\% level to mitigate the influence of extreme values.

Table \ref{tab:cycles} presents the results. Liquidity updates on decentralized exchanges are very infrequent, as times elapsed between consecutive interactions are measured in days or even weeks. In line with Prediction \ref{pred:updates}, we find evidence for shorter liquidity cycles on low-fee pools. The average time between consecutive mint and burn orders is 22.05\% shorter on the low-fee pool (from Model 2, the relative difference is 112.42 hours/509.19 hours).


We repeat the analysis above with burn-to-mint times as the dependent variables. The burn-to-mint time measures the speed at which $\LP$s deposit liquidity at updated prices after removing (out-of-range) positions. Our findings reveal that \(\LP\)s in low-fee pools replenish liquidity 63\% faster than those in high-fee pools. This supports the notion that \(\LP\)s in low-fee environments are larger, more sophisticated market participants.  

\begin{table}[H]
\caption{Liquidity cycles on fragmented pools} \label{tab:cycles}
\begin{minipage}[t]{1\columnwidth}%
\footnotesize
			This table reports the coefficients of the following regression:
	\begin{align*}
    y_{ijtk}=\alpha + \beta_0 d_\emph{low-fee, ij} + \beta_1 \text{GasPrice}_{jt} d_\emph{low-fee, ij} + \beta_2 \text{GasPrice}_{jt} \times d_\emph{high-fee, ij} + \sum \beta_k \emph{Controls}_{ijt} + \theta_j + \varepsilon_{ijt}
    \end{align*}
	where the dependent variable $y_{ijt}$ can be (i) the mint-to-burn time, (ii) the burn-to-mint time, measured in hours, for a transaction initiated by wallet $k$ on day $t$ and pool $i$ trading asset $j$. The mint-to-burn and burn-to-mint times are computed for consecutive interactions of the same wallet address with the liquidity pool. $d_\emph{low-fee, ij}$ is a dummy that takes the value one for the pool with the lowest fee in pair $j$ and zero else. $d_\emph{high-fee, ij}$ is defined as $1-d_\emph{low-fee, ij}$. $\emph{GasPrice}_{jt}$ is the average of the lowest 100 bids on liquidity provision events across all pairs on day $t$, standardized to have a zero mean and unit variance. \emph{Volume} is the natural logarithm of the sum of all swap amounts on day $t$, expressed in thousands of US dollars. \emph{Total value locked} is the natural logarithm of the total value locked on Uniswap v3 pools on day $t$, expressed in millions of dollars. \emph{Volatility} is computed as the daily range between high and low prices on the most active pool for a given pair. \emph{Position out-of-range} is a dummy taking value one if the position being burned or minted is out of range, that is if the price range selected by the $\LP$ does not straddle the current pool price. All variables are measured as of the time of the second leg of the cycle (i.e., the burn of a mint-burn cycle). All regressions include pair, week, and trader wallet fixed-effects. Robust standard errors in parenthesis are clustered by day and ***, **, and * denote the statistical significance at the 1, 5, and 10\% level, respectively.  The sample period is from May 4, 2021 to July 15, 2023. 
\end{minipage}
\begin{center}
\begin{tabular}{lcccccc}
\toprule
 & \multicolumn{4}{c}{Mint-burn time (hours)} & \multicolumn{2}{c}{Burn-mint time (hours)}  \\
 & \multicolumn{2}{c}{Out-of-range positions} & \multicolumn{2}{c}{Full sample} \\
 & (1) & (2) & (3) & (4) & (5) & (6)  \\
\cmidrule{1-7}
$ d_\text{low-fee}$ & -110.94*** & -112.42*** & -99.74*** & -100.17*** & -157.95*** & -159.71*** \\
 & (-7.49) & (-7.69) & (-8.86) & (-8.94) & (-10.59) & (-10.81) \\
Gas price $\times$ $ d_\text{low-fee}$ & -14.27 & -6.54 & -16.65** & -15.41* & -11.29 & 2.95 \\
 & (-1.49) & (-0.68) & (-2.13) & (-1.98) & (-1.65) & (0.40) \\
Gas price $\times$ $ d_\text{high-fee}$ & -19.57** & -12.83 & -14.44** & -13.42* & -10.52* & 1.96 \\
 & (-2.34) & (-1.57) & (-2.04) & (-1.89) & (-1.69) & (0.32) \\
Volume &  & -16.71*** &  & -5.87 &  & -24.84*** \\
 &  & (-3.24) &  & (-1.15) &  & (-4.10) \\
Total value locked &  & -35.14 &  & -53.17* &  & -12.71 \\
 &  & (-1.05) &  & (-1.70) &  & (-0.52) \\
Volatility &  & -3.48** &  & -2.11*** &  & -2.99*** \\
 &  & (-2.49) &  & (-2.75) &  & (-3.36) \\
Constant & 509.19*** & 509.66*** & 497.18*** & 497.00*** & 248.00*** & 250.13*** \\
 & (61.93) & (58.34) & (91.65) & (90.60) & (29.91) & (30.27) \\
 &  &  &  &  &  &  \\
 Pair FE & Yes & Yes & Yes & Yes & Yes & Yes  \\
 Week FE & Yes & Yes & Yes & Yes & Yes & Yes  \\
 Trader wallet FE & Yes & Yes & Yes & Yes & Yes & Yes \\
Observations & 215,454 & 215,454 & 405,586 & 405,584 & 265,848 & 265,848 \\
 R-squared & 0.87 & 0.87 & 0.82 & 0.82 & 0.37 & 0.37 \\ \hline
\bottomrule
\multicolumn{7}{l}{Robust t-statistics in parentheses. Standard errors are clustered at week level.} \\
\multicolumn{7}{l}{*** p$<$0.01, ** p$<$0.05, * p$<$0.1} \\
\end{tabular}

\end{center}
\end{table}

\subsection{Adverse selection costs across low- and high-fee pools}

Finally, we test Prediction \ref{pred:as} of our model, which states that $\LP$ on the low-fee pool face higher adverse selection costs. Our main metric for informational costs is the \emph{loss-versus-rebalancing} (LVR), as defined in \citet{zhang2023amm}. The measure is equivalent to the adverse selection component of the bid-ask spread in equity markets. To calculate it, for each swap \(j\) exchanging \(\Delta x_j\) for \(\Delta y_j\) in a pool with assets \(x\) and \(y\), we use:
\begin{equation}
      \text{LVR}_{j}=d_j \times \Delta x_j (p_{\text{swap},j}-p^\prime_j),
\end{equation}
where \(d_j\) is one for a ``buy'' trade (\(\Delta x_j<0\)) and minus one for a ``sell'' trade (\(\Delta x_j>0\)). The effective swap price is \(p_{\text{swap},j}=-\frac{\Delta y_j}{\Delta x_j}\), and \(p^\prime_j\) represents a benchmark price.

We use two benchmark prices \(p^\prime_j\) in our analysis. The first, \(p^\prime_j=p_{j}^{\Delta t=0}\), is the pool's equilibrium price immediately after a swap. The resulting LVR metric captures both temporary and permanent price impact, driven by uninformed and informed trades, respectively, and represents an upper bound for $\LP$'s adverse selection cost. 

The second benchmark is the liquidity-weighted average price across Uniswap v3 pools, measured with a one-hour delay after the swap (\(p^\prime_j=p_{j}^{\Delta t=1h}\)). This approach assumes that any price deviations caused by uninformed liquidity trades are corrected by arbitrageurs within an hour, as supported by \citet{LeharParlour2021}. Thus, the LVR metric derived using this benchmark captures only the permanent price impact, a more precise measure of adverse selection cost for liquidity providers.\footnote{Our methodology is equivalent to the one in \citet{zhang2023amm} under two assumptions. First, liquidity providers can re-balance their position following each swap. Second, we assume that our two benchmarks for $p^\prime_j$, derived from decentralized exchange data, closely track the fundamental value of the token. This perspective aligns with \citet{han2022trust}, who also note that centralized exchange prices are subject to manipulative practices such as wash trading. Further, our selection of benchmarks reflects the fact that our sample includes several token pairs not traded on major centralized exchanges such as Binance.}

To compute LVR for each day $t$ and liquidity pool $i$, we aggregate the loss-versus-balancing for each swap within a day. We subsequently winsorize our measures at the 0.5\% and 99.5\% quantiles to remove extreme outliers. The resulting sum is normalized by dividing it by the total value locked (TVL) in the pool at day's end:
\begin{equation}\label{eq:LVR}
      \text{LVR}_{i,t}=\frac{\sum_j \text{LVR}_{j, i,t}}{\text{TVL}_{i,t}},  
\end{equation}
which ensures that the LVR metric is comparable across pools trading different token pairs.

We complement our analysis with the calculation of \emph{impermanent loss} (IL), an additional metric for assessing adverse selection costs. Impermanent loss is defined as the negative return from providing liquidity compared to simply holding the assets outside the exchange and marking them to market as prices change \citep[see, for example,][]{aoyagi2020,BarbonRanaldo2021}.

The key distinction between IL and loss-versus-rebalancing (LVR) measures lies in their assumptions about liquidity providers' strategies \citep{zhang2023amm}. While LVR assumes that providers actively re-balance their holdings by mirroring decentralized exchange trades on centralized exchanges at the fundamental value to hedge market risk, IL is based on a more passive approach where providers maintain their positions without active re-balancing. Loss-versus-rebalancing is a function of the entire price path, reflecting constant rebalancing by liquidity providers. In contrast, impermanent loss is determined solely by the initial and final prices of the assets.

We obtain hourly liquidity snapshots from the Uniswap V3 Subgraph to calculate impermanent loss for a theoretical symmetric liquidity position. This position is set within a price range of \(\left[\frac{1}{\alpha} p, \alpha p\right]\), centered around the current pool price \(p\), with \(\alpha\) set to 1.05. We set a one-hour horizon to measure changes in position value, aligning with the time horizon used for the LVR metric. In Appendix \ref{app:IL}, we present the exact formulas for calculating impermanent loss on Uniswap V3, based on the methodology described by \citet{Heimbach2023}.

\begin{table}
\scriptsize
\caption{Adverse selection costs on high- and low-fee pools}\label{tab:lvr}
\begin{minipage}[t]{1\columnwidth}%
\footnotesize
This table presents regression results that analyze adverse selection costs in fragmented Uniswap v3 pools. For columns (1) through (4), the dependent variable is loss-versus-rebalancing (LVR), as defined in equation \eqref{eq:LVR}. We use the one-hour horizon benchmark (\(p_j^{\Delta t=1h}\)) in models (1) and (2) to measure permanent price impact, and the immediate, same-block price benchmark (\(p_j^{\Delta t=0}\)) in models (3) and (4) to measure total price impact. For columns (5) and (6), the dependent variable is the impermanent loss for a symmetric liquidity position at $\pm 5\%$ centered around the current pool price. The average impermanent loss is calculated for each day, based on Ethereum blocks mined within that day. The impermanent loss computation uses a one-hour liquidity provider horizon, comparing current pool prices with those one hour later. For columns (7) and (8), the dependent variables are the liquidity (TVL) and volume share of the pool, measured in percent. Finally, in columns (9) and (10) the dependent variable is the absolute deviation of the Uniswap pool price from Binance prices, sampled hourly, and measured in percent.
$d_\emph{low-fee, ij}$ is a dummy that takes the value one for the pool with the lowest fee in pair $j$ and zero else. $\emph{GasPrice}_{jt}$ is the average of the lowest 100 bids on liquidity provision events across all pairs on day $t$, standardized to have a zero mean and unit variance.  \emph{Volume} is the natural logarithm of the sum of all swap amounts on day $t$, expressed in thousands of US dollars. \emph{Total value locked} is the natural logarithm of the total value locked on Uniswap v3 pools on day $t$, expressed in millions of dollars.  When LVR is an explanatory variable, it is calculated using the one-hour ahead benchmark price.  \emph{Volatility} is computed as the daily range between high and low prices on the most active pool for a given pair. All regressions include pair and week fixed-effects. Robust standard errors in parenthesis are clustered by week, and ***, **, and * denote the statistical significance at the 1, 5, and 10\% level, respectively.  The sample period is from May 4, 2021 to July 15, 2023. 
\end{minipage}

\begin{center}
\resizebox{1\textwidth}{!}{  
\begin{tabular}{lcccccccc}
\toprule
& \multicolumn{2}{c}{LVR (1h horizon)} & \multicolumn{2}{c}{LVR (after swap)} & \multicolumn{2}{c}{Impermanent loss} & \multicolumn{2}{c}{CEX price deviation} \\
& \multicolumn{2}{c}{Permanent price impact} & \multicolumn{2}{c}{Total price impact} \\
\cmidrule{1-9} 
 & (1) & (2) & (3) & (4) & (5) & (6) & (7) & (8)  \\
\cmidrule{1-9}
$ d_\text{low-fee}$ & 6.39*** & 6.39*** & 29.78*** & 29.67*** & 1.08*** & 1.13*** & 0.06 & 0.04 \\
 & (16.57) & (17.05) & (14.86) & (14.95) & (5.72) & (6.18) & (1.51) & (1.33) \\
Gas price $\times$ $ d_\text{low-fee}$ &  & -0.75** &  & 3.51** &  & -0.01 &  & 0.08 \\
 &  & (-2.05) &  & (2.10) &  & (-0.05) &  & (1.09) \\
Gas price &  & 2.61*** &  & 6.16** &  & 3.71*** &  & -0.03 \\
 &  & (2.74) &  & (2.53) &  & (3.76) &  & (-0.28) \\
Volume &  & 3.22*** &  & 8.67*** &  & 1.81*** &  & 0.22*** \\
 &  & (8.15) &  & (6.61) &  & (6.22) &  & (4.74) \\
Total value locked &  & 0.53 &  & -2.12 &  & 1.93 &  & -0.39*** \\
 &  & (0.14) &  & (-0.34) &  & (0.74) &  & (-4.05) \\
Volatility &  & 1.85*** &  & 4.23*** &  & 6.69** &  & 1.04*** \\
 &  & (2.87) &  & (3.17) &  & (2.61) &  & (3.51) \\
Constant & 7.85*** & 7.86*** & 8.88*** & 8.97*** & 7.37*** & 7.51*** & 0.60*** & 0.67*** \\
 & (40.71) & (36.89) & (8.87) & (8.88) & (77.84) & (61.32) & (30.71) & (35.75) \\
 &  &  &  &  &  &  \\
Pair FE & Yes & Yes & Yes & Yes & Yes & Yes & Yes & Yes \\
Week FE & Yes & Yes & Yes & Yes & Yes & Yes & Yes & Yes \\
Observations & 40,302 & 40,288 & 40,302 & 40,288 & 40,250 & 40,248 & 5,207 & 5,207 \\
 R-squared & 0.14 & 0.15 & 0.09 & 0.10 & 0.09 & 0.11 & 0.10 & 0.11 \\ \hline
\bottomrule
\multicolumn{9}{l}{Robust t-statistics in parentheses. Standard errors are clustered at week level.} \\
\multicolumn{9}{l}{*** p$<$0.01, ** p$<$0.05, * p$<$0.1} \\
\end{tabular}
}
\end{center}
\end{table}

Table \ref{tab:lvr} presents our empirical results. In line with \citet{milionis2023automated}, all price impact measures --- that is, the immediate and one-hour horizon LVR and the impermanent loss --- are significantly larger in low-fee pools. This indicates that a higher liquidity fee indeed acts as barrier to arbitrageurs. Specifically, the permanent price impact, measured by the one-hour horizon LVR, is 6.39 basis points or 81\% larger in low-fee pools. The total price impact, represented by the after-swap LVR metric, is 3.5 times larger in low-fee pools. The wide gap between permanent and total price impact highlights the substantially higher volume of uninformed trading in low-fee pools. Our secondary measure of adverse selection, the impermanent loss at 5\% around the current price, is also 15\% higher on low- than high-fee pools.  

We note that an increase in gas prices leads to a 3.51 wider gap in total price impact but a 0.75 bps narrower gap in permanent price impact between the high- and low-fee pools. The result suggests that a higher gas price primarily discourages uninformed traders, rather than arbitrageurs, from trading on high-fee pools. 


Finally, in Models (7) and (8) of Table \ref{tab:lvr} we explore whether various arbitrage frictions lead to price discrepancies between high- and low-fee pools. For this purpose, we collect hourly price data from Binance for the largest four pairs by trading volume: WBTC-WETH, USDC-WETH, WETH-USDT, and USDT-USDC. We subsequently compute daily averages of hourly price deviations between centralized and decentralized exchanges. The analysis reveals that the average hourly price deviation across centralized and decentralized exchanges is 0.60\%. Notably, there is no significant difference in price deviations between low- and high-fee pools. The result suggests that arbitrage activities remain efficient despite the differences in trading costs between these pools.

\begin{figure}[H]
\caption{\label{fig:lvr} Price impact and price deviations across high- and low-fee pools}
\begin{minipage}[t]{1\columnwidth}%
\footnotesize
This figure plots the average total and permanent price impact, liquidity yield, and price deviation from centralized exchanges across low and high fee pools for fragmented pairs.
\end{minipage}

\vspace{0.05in}

\begin{centering}

\includegraphics[width=\textwidth]{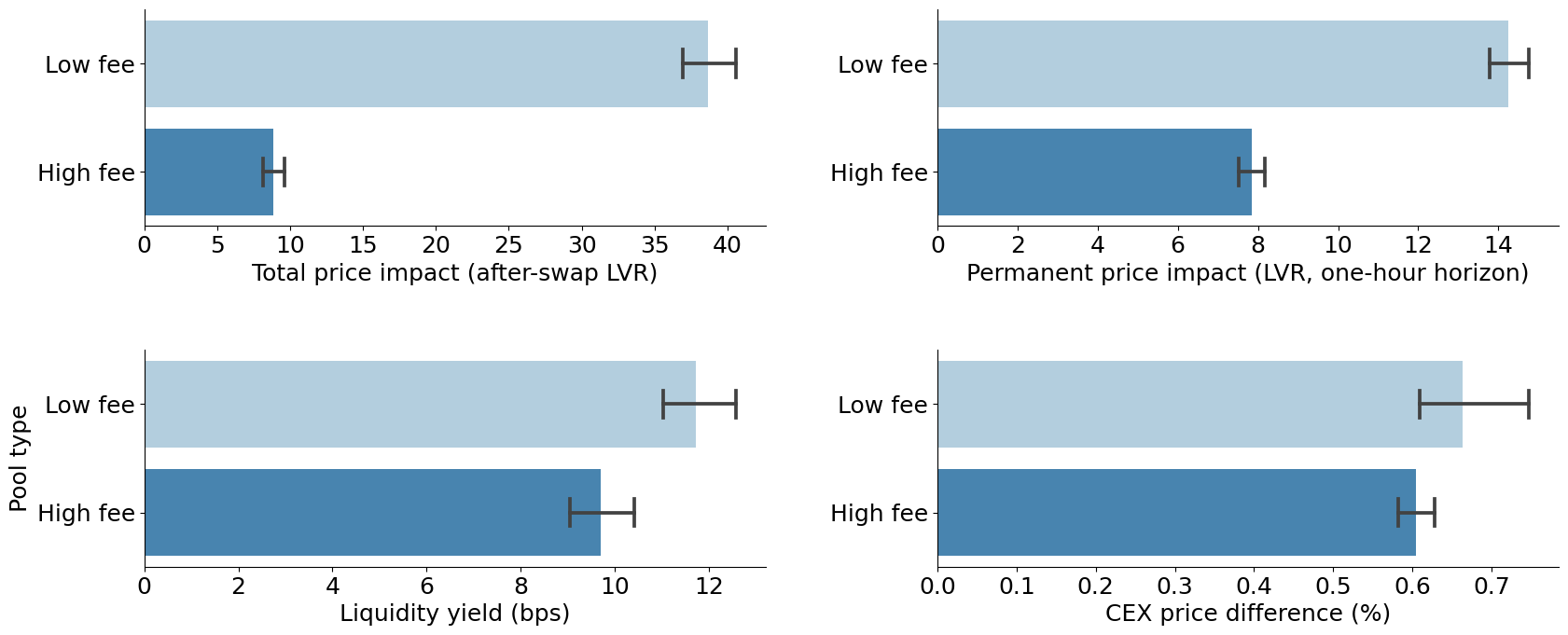}

\par\end{centering}

\end{figure}

Figure \ref{fig:lvr} graphically illustrates the result, contrasting permanent price impact measures against the liquidity yield, as calculated in equation \eqref{eq:liq_yield}. Notably, before accounting for gas costs, we observe that liquidity providers in low-fee pools experience losses on average: the average daily permanent price impact in these pools is 14.21 basis points, which exceeds the fee revenue of 11.71 bps.  In contrast, liquidity providers in high-fee pools approximately break even before considering gas costs: the fee revenue amounts to 9.69 bps, which is slightly higher than the permanent price impact of 7.85 bps. One should keep in mind, however, that the magnitude of losses from adverse selection depends on the horizon at which we measure the loss-versus-rebalancing.

\section{Conclusion}
This paper argues that fixed costs associated with liquidity management drive a wedge between large (institutional) and small (retail) market makers. In the context of blockchain-based decentralized exchanges, the most evident fixed cost is represented by gas fees, where market makers compensate miners and validators for transaction processing in proof-of-work, respectively in proof-of-stake blockchains. Innovative solutions such as Proof of Stake (PoS) consensus algorithms and Layer 2 scaling aim to address the concern of network costs. However, even if gas fees were eliminated entirely, individual retail traders still encounter disproportionate fixed costs in managing their liquidity, such as the expenditure of time and effort.

Our paper highlights a trade-off between capital efficiency and the fixed costs of active management. During the initial phase of decentralized exchanges, such as Uniswap V2, liquidity providers were not able to set price limits, resulting in an even more passive liquidity supply and fewer incentives for active position management. However, the mechanism implied that incoming trades incurred significant price impact. To enhance the return on liquidity provision and reduce price impact on incoming trades, modern decentralized exchanges (DEXs) have evolved to enable market makers to fine-tune their liquidity positions, albeit at the expense of more active management.

We show, both theoretically and empirically, that fixed costs of liquidity management promote market fragmentation across decentralized pools and generate clienteles of liquidity providers. Large market makers, likely institutions and funds, have stronger economies of scale and can afford to frequently manage their positions on very active low fee markets, while bearing higher adverse selection risk. On the other hand, smaller retail liquidity providers become confined to high fee markets with scant activity, trading off a lower execution probability against reduced adverse selection and lower gas costs to update their positions. Since large liquidity providers can churn their position at a faster pace, two thirds of the trading volume interacts with less than half the capital locked on Uniswap V3.

Our findings indicate that substantial fixed costs can hinder the participation of small market makers in the forefront of liquidity provision, where active order management is crucial. Instead, smaller liquidity providers tend to operate on the market maker ``fringe,'' opting for a lower execution probability in exchange for better prices. The results are particularly relevant the context of a resurgence in retail trading activity and the ongoing evolution of technology that fosters market structures aimed at enhancing broader access to financial markets.

\bibliographystyle{jf}
\bibliography{references}

\newpage

\appendix

\numberwithin{equation}{section}
\numberwithin{prop}{section}
\numberwithin{lem}{section}
\numberwithin{defn}{section}
\numberwithin{cor}{section}
\numberwithin{figure}{section}
\numberwithin{table}{section}

\let\normalsize\small

\appendix

\section{Liquidity provision mechanism on Uniswap v3 \label{sec:app-dex}}
In this appendix, we walk through a numerical example to illustrate the mechanism of liquidity provision and trading on Uniswap V3 liquidity pools. To facilitate understanding, we highlight the similarities and differences between the Uniswap mechanism and the familiar economics of a traditional limit order book.

\begin{figure}[H]
\caption{\label{fig:app_grid} ETH/USDT price grid around $p_c$}
\begin{center}
    \includegraphics[width=0.75\textwidth]{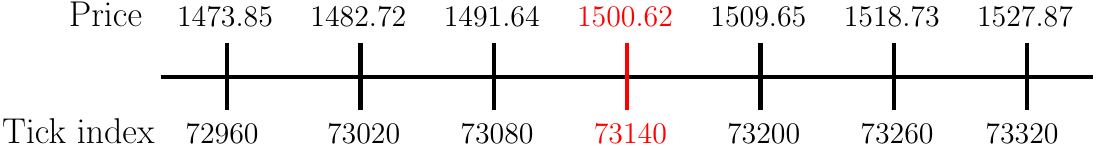}
\end{center}
\end{figure}

Let $p_c=1500.62$ be the current price of the ETH/USDT pair. Traders can provide liquidity on Uniswap V3 pools at prices on a log-linear tick space. In particular, consecutive prices are always $\theta$ basis point apart: $p_i=1.0001^{\theta i}$, where $\theta$ is the tick spacing. For the purpose of the example, we take $\theta=60$. Consequently, the current price of $1500.62$ corresponds to a tick index of $c=73140$. Figure \ref{fig:app_grid} illustrates three ticks on grid below and above the current price of ETH/USDT $1500.62$.

\paragraph{Two-sided liquidity provision.} Trader \textbf{A} starts out with a capital of USDT 20,000 and wants to provide liquidity over the price range $\left[1491.64, 1527.87\right]$, a range which spans four ticks. Liquidity provision over a range that includes the current price corresponds to posting quotes on both the bid and ask side of a traditional limit order book, where the current price of the pool corresponds to the mid-point of the book.
\begin{enumerate}
    \item \emph{Bid quotes:} trader \textbf{A} deposits USDT over the price range $\left[1491.64,1500.62\right)$. This action is equivalent to submitting a buy limit order with a bid price of $1491.64$. An incoming Ether seller can swap their ETH for the USDT deposited by \textbf{A}, generating price impact until the limit price of $1491.64$ is reached.
    \item \emph{Ask quotes:} at the same time, trader \textbf{A} deposits ETH over three ticks: $\left[1500.62,1509.65\right)$, $\left[1509.65,1518.73\right)$, and $\left[1518.73,1527.87\right)$. The action corresponds to submitting \emph{three} sell limit orders with ask prices $1509.65$, $1518.73$, and $1527.87$, respectively. Incoming Ether buyers can swap USDT for trader \textbf{A}'s ETH.
\end{enumerate}

In the Uniswap V3 protocol, deposit amounts over each tick $\left[p_i,p_{i+1}\right)$ must satisfy
\begin{align}\label{eq:app_deposits_tick}
    \text{ETH deposit over $\left[p_{i},p_{i+1}\right)$: } x_i&=L\left(\frac{1}{\sqrt{p_i}}-\frac{1}{\sqrt{p_{i+1}}}\right) \\
    \text{USDT deposit over $\left[p_{i},p_{i+1}\right)$: } y_i &= L\left(\sqrt{p_{i+1}}-\sqrt{p_i}\right), 
\end{align}
where $L$ (``liquidity units'') is a scaling factor proportional to the capital committed to the liquidity position. The scaling factor $L$ is pinned down by setting the total committed capital equal to the sum of the positions (in USDT), that is  $p_c\sum_{i} x_i + \sum_{i} y_i$. In our example,
\begin{equation}
    1500.62 \times L_A \times \left(\frac{1}{\sqrt{1500.62}}-\frac{1}{\sqrt{1527.87}}\right) + L_A \times \left(\sqrt{1500.62}-\sqrt{1491.64}\right) = 20000,
\end{equation}
leading to $L_A=43188.6$. We the value of $L_A$ into \eqref{eq:app_deposits_tick} and conclude that trader \textbf{A} deposits 5,013.38 USDT over $\left[1491.64, 1500.62\right)$ and ETH 9.99 over  $\left[1500.62, 1527.87\right)$ (approximately ETH 3.33 over each tick size covered).

\paragraph{One-sided liquidity provision.} Trader \textbf{B} has USDT 20,000 and wants to post liquidity over the range $\left[1509.65,1527.87\right)$, which does not include the current price.  This action corresponds to posting ask quotes to sell ETH deep in the book, at price levels $1518.73$ and $1527.87$. Liquidity is not ``active'' -- that is, the quotes are not filled -- until the existing depth at $1509.65$ is consumed by incoming trades. 

We use equation \eqref{eq:app_deposits_tick} to solve for the amount of liquidity units provided by \textbf{B}:
\begin{equation}
    1500.62 \times L_B \times \left(\frac{1}{\sqrt{1509.65}}-\frac{1}{\sqrt{1527.87}}\right) = 20000,
\end{equation}
which leads to $L_B=86589.4$. Trader \textbf{B} deposits 6.67 ETH on each of the two ticks covered by the chosen range. 

\begin{figure}[H]
\caption{\label{fig:liquidity_pool} ETH/USDT pool state after liquidity provision choices}
\begin{center}
    \includegraphics[width=\textwidth]{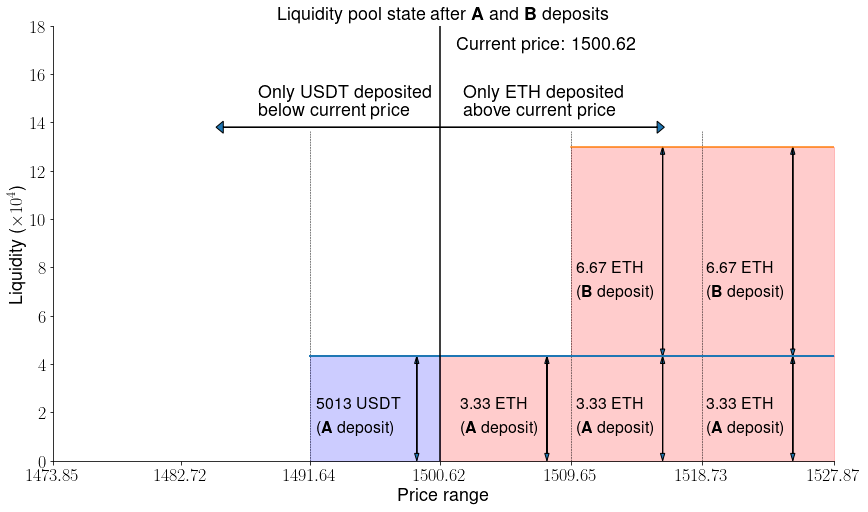}
\end{center}
\end{figure}

Figure \ref{fig:liquidity_pool} illustrates market depth after \textbf{A} and \textbf{B} deposit liquidity in the pool. The current price of the pool is equivalent to a midpoint in traditional limit order markets. The ``ask side'' of the pool is deeper, consistent with both liquidity providers choosing ranges skewed towards prices above the current midpoint. Liquidity is uniformly provided over ticks -- that is, each trader deposits an equal share of their capital at each price tick covered by their price range.

\paragraph{Trading, fees, and price impact.} Suppose now that a trader \textbf{C} wants to buy 10 ETH from the pool. For each tick interval $\left[p_i,p_{i+1}\right)$, price impact is computed using a constant product function over virtual reserves:
\begin{equation}
    \underbrace{\left(x+\frac{L}{\sqrt{p_{i+1}}}\right)}_\text{Virtual ETH reserves}\underbrace{\left(y+L\sqrt{p_i}\right)}_\text{Virtual USDT reserves}=L^2,
\end{equation}
where $x$ and $y$ are the actual ETH and USDT deposits in that tick range, respectively. Virtual reserves are just a mathematical artifact: they extend the physical (real) deposits as if liquidity would be uniformly distributed over all possible prices on the real line. Working with constant product functions over real reserves is not feasible: in our example, the product of real reserves is zero throughout the order book (since only one asset is deposited in each tick range).

Let $\tau=1\%$ denote the pool fee that serves as an additional compensation for liquidity providers. That is, if the buyer pays to pay $\Delta y$ USDT to purchase a quantity $\Delta x$ ETH, he needs to effectively pay $\Delta y\left(1+\tau\right)$. As per the Uniswap V3 white paper, liquidity fees are not automatically deposited back into the pool.

\begin{enumerate}
    \item \textbf{Tick 1: $\left[1500.62,1509.65\right)$}. Trader \textbf{C} first purchases $3.33$ ETH at the first available tick above the current price (equivalent to the ``best ask''). To remove the ETH, he needs to deposit $\Delta y_1$ USDT, where $\Delta y_1$ solves:
    \begin{equation}
        \left(3.33-3.33+\frac{L_A}{\sqrt{1509.65}}\right)\left(0+\Delta y_1+L_A\sqrt{1500.62}\right)=L_A^2,
    \end{equation}

which leads to $\Delta y_1=5026.19$ USDT. Trader \textbf{C} pays an average price of 50216.19/3.33=1507.86 USDT for each unit of ETH purchased. Further, he pays a fee of 50.26 USDT to liquidity provider \textbf{A} (the only liquidity provider at this tick).

The new current price is given by the ratio of virtual reserves,
\begin{equation}
    p^\prime=\frac{\Delta y_1 +L_A\sqrt{1500.62}}{3.33-3.33+\frac{L_A}{\sqrt{1509.65}}}=1509.65,
\end{equation}
that is the next price on the tick grid since \textbf{C} exhausts the entire liquidity on $\left[1500.62, 1509.65\right)$.

\item \textbf{Tick 2: $\left[1509.65,1518.73\right)$}. Trader \textbf{C} still needs to purchase 6.67 ETH at the next tick level (where the depth is 10 ETH). The liquidity level at this tick is $L_A+L_B$, that is the sum of liquidity provided by \textbf{A} and \textbf{B}. To remove the 6.67 ETH from the pool, he needs to deposit $\Delta y_2$, where
\begin{equation}
    \left(10-6.67+\frac{L_A+L_B}{\sqrt{1518.73}}\right)\left(0+\Delta y_2 + \left(L_A+L_B\right)\sqrt{1509.65}\right)=\left(L_A+L_B\right)^2.
\end{equation}
It follows that trader \textbf{C} purchases 6.67 ETH by depositing $\Delta y_2=10089.12$ USDT, at an average price of 1512.61. The pool price is updated as the ratio of virtual reserves:
\begin{equation}
    p^{\prime\prime}=\frac{\Delta y_2 +\left(L_A+L_B\right)\sqrt{1509.65}}{10-6.67+\frac{L_A+L_B}{\sqrt{1518.73}}}=1515.7.
\end{equation}
The updated price is in between the two liquidity ticks, since not all depth on this tick level was exhausted in the trade. Following the swap, liquidity on the tick range $\left[1509.65,1518.73\right)$ is composed of both assets: that is 10089.12 USDT and 10-6.67=3.33 ETH. 

Finally, trader $C$ pays 100.89 USDT as liquidity fees (1\% of the trade size), which are distributed to \textbf{A} and \textbf{B} proportionally to their liquidity share. That is, \textbf{A} receives a fraction $\frac{L_A}{L_A+L_B}$ of the total fee (33.57 USDT), whereas \textbf{B} receives 67.32 USDT.

\end{enumerate}

\begin{figure}[H]
\caption{\label{fig:swap_pool} Swap execution and price impact}
\begin{center}
    \includegraphics[width=\textwidth]{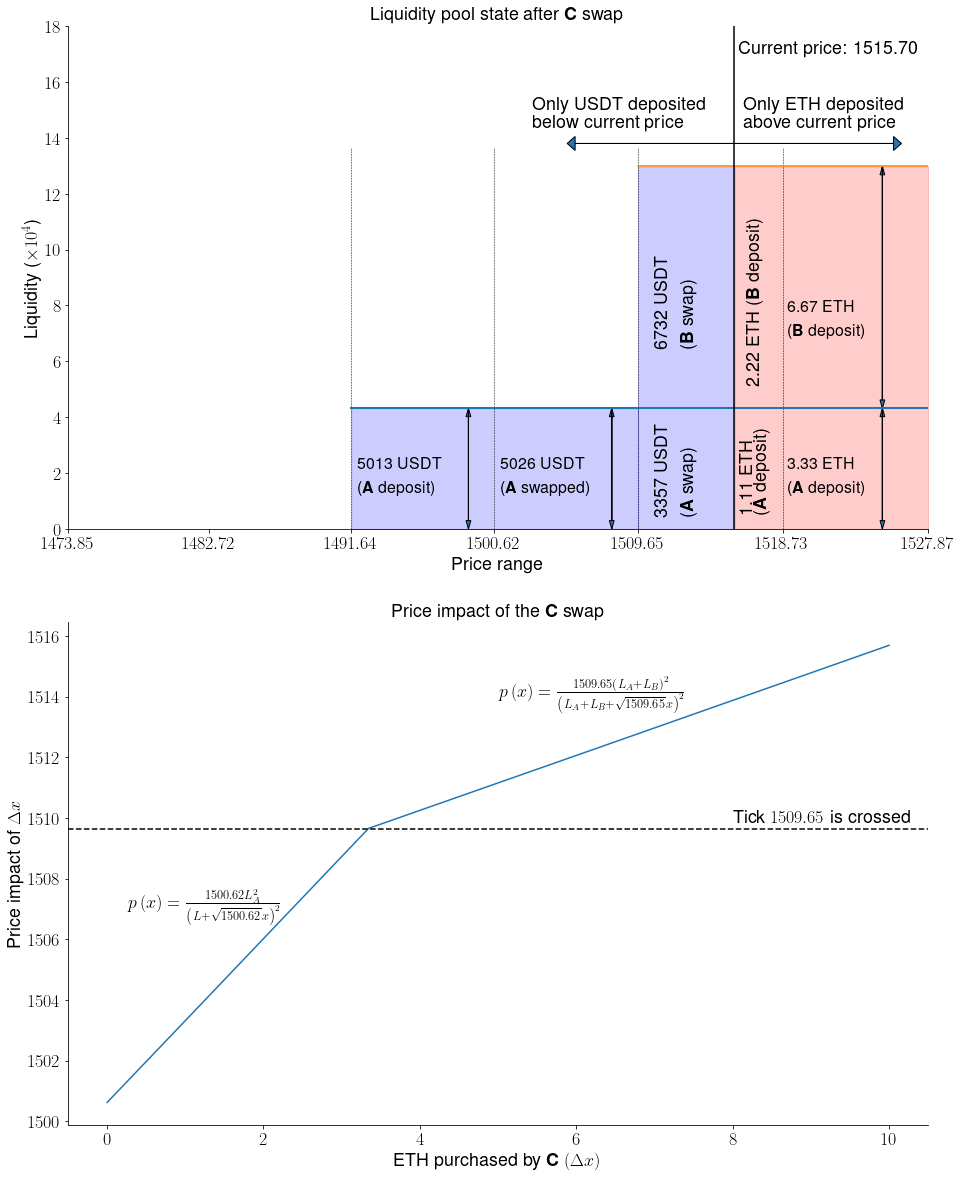}
\end{center}
\end{figure}

Figure \ref{fig:swap_pool} illustrates the impact of the swap. Within tick $\left[1500.62,1509.65\right)$, \textbf{A} sells 3.33 ETH and buys 5026 USDT. Unlike on limit order books, the execution does not remove liquidity from the book. Rather, $\textbf{A}$'s capital is converted from one token to another and remains available to trade. This feature underscores the passive nature of liquidity supply on decentralized exchanges. Mapping the concepts to traditional limit order book, this mechanism would imply that every time a market maker's sell order is executed at the ask, a buy order would automatically be placed on the bid side of the market.

The final price of $1517.70$ lies within the tick $\left[1509.65,1518.73\right)$, rather than on its boundary. Trader \textbf{C} only purchases 6.66 ETH out of 10 ETH available within this price interval. The implication is that liquidity on $\left[1509.65,1518.73\right)$ contains both tokens: 3.33 ETH (the amount that was not swapped by \textbf{C}) as well as 10089.12 USDT that \textbf{C} deposited in the pool. 

The bottom panel of Figure \ref{fig:swap_pool} shows the price impact of the swap. From equation (6.15) in the Uniswap V3 white paper, we can solve for the price within tick $\left[p_\text{min},p_\text{max}\right)$ with liquidity $L$, following the execution of a buy order of size $x$:
\begin{equation}
    p\left(x\right)=\frac{p_\text{min} L^2}{\left(L-\sqrt{p_\text{min}}x\right)^2}.
\end{equation}
As expected, the price impact of a swap decreases in the liquidity available $L$ -- each ETH unit purchased by \textbf{C} has a smaller impact on the price once tick 1509.65 is crossed and the market becomes deeper.

\newpage

\section{Notation summary\label{sec:Variable-Definitions}}

\onehalfspacing
\noindent
\begin{center}
\begin{tabular}{@{}ll@{}}
\toprule
\cmidrule{1-2}

\multicolumn{2}{c}{\textbf{Variable Subscripts}}\\
\cmidrule{1-2}
Subscript & Definition \\
\cmidrule{1-2}
\textbf{T} and \textbf{N} & Pertaining to the token and numeraire assets, respectively. \\
\textbf{L} and \textbf{H} & Pertaining to the low- and high-fee pool, respectively. \\
\textbf{LP} & Pertaining to liquidity providers. \\
\textbf{LT} & Pertaining to liquidity traders. \\
\textbf{A} & Pertaining to arbitrageurs. \\
\cmidrule{1-2}

\multicolumn{2}{c}{\textbf{Exogenous Parameters}}\\
\cmidrule{1-2}
Parameters & Definition \\
\cmidrule{1-2}
$v_t$ & Common value of the token at time $t$. \\
$\eta$, $1-\eta$ & Poisson arrival rate of news and private value shocks, respectively. \\
$\delta$ & Size of common or private value shock. \\
$\Delta$ & Parameter governing the probability distribution of shocks, $ \phi\left(\delta\right)=\frac{1}{2\Delta \sqrt{1+\delta}}$. \\
$\ell$, $h$ & Liquidity fee on the low- and high-fee pool. \\
$f$ & Liquidity fee on a non-fragmented pool. \\
$q_i$ & Token endowment of liquidity provider $i$, exponentially distributed with scale parameter $\lambda$. \\
$\lambda$ & Aggregate liquidity supply if all $\LP$s join the market. \\
$\Gamma$ & Gas price on the blockchain. \\
$r$ & Width of the price range $\left[\frac{v}{\left(1+r\right)^2}, v(1+r)^2\right]$. \\
\cmidrule{1-2}

\multicolumn{2}{c}{\textbf{Endogenous Quantities}}\\
\cmidrule{1-2}
Variable & Definition \\
\cmidrule{1-2}
$T_k$ & Equilibrium liquidity supply on exchange $k\in\left\{\textbf{L},\textbf{H}\right\}$. \\
$\tau\left(\delta\right)$ & Optimal trade size for $\LT$ or $\A$ with value shock $\delta$. \\
$\mathcal{L}\left(f_k\right)$ & Liquidity yield (fee revenue per unit of liquidity supplied) on pool with fee $f_k$. \\
$\mathcal{A}\left(f_k\right)$ & Adverse selection cost for $\LP$ per unit of liquidity supplied on pool with fee $f_k$. \\
$\mathcal{C}\left(f_k\right)$ & Liquidity re-balancing cost on pool with fee $f_k$. \\
$\qmg^\star$ & Token endowment of the $\LP$ who is indifferent between pools. \\
$\underline{q}_k$ &  Lowest token endowment deposited on pool $k$ (from break-even condition). \\
$\pi_k$ & Expected liquidity provider profit on exchange $k$. \\
$w_k$ & Liquidity market share for pool with fee $k$. \\

\bottomrule
\end{tabular}
\end{center}

\newpage

\section{Proofs \label{sec:proofs}}
\noindent \textbf{\large Lemma \ref{lem:liq_revenue}}
\begin{proof}
We take the expectation of fee revenues over the size of private value shocks $\delta$ and obtain:
\begin{align}
    \mathbb{E}\text{ProfitLiq}_{i,k}&=2q_i v f_k \times \Big\{ \nonumber \\ &\mathbb{P}\left(f_k<\delta \leq (1+f_k)(1+r)^2-1\right) \times \frac{1+r}{r}\mathbb{E} \left[\sqrt{\frac{1+\delta}{1+f_k}}-1 \mid f_k<\delta \leq (1+f_k)(1+r)^2-1 \right] + \nonumber \\
    &+ \mathbb{P}\left(\delta>(1+f_k)(1+r)^2-1\right) \times \left(1+r\right) \Big\} \nonumber \\
    &=q_i \underbrace{v \frac{f_k (r+1) \left(2 \Delta -r\sqrt{f_k+1} -2 \sqrt{f_k+1}\right)}{\Delta }}_{\equiv \mathcal{L}\left(f_k\right)},
\end{align}
where we define $\mathcal{L}\left(f_k\right)$ as the liquidity yield: i.e., the per-unit profit from liquidity provision in pool $k$.

To explore how the liquidity revenue changes with respect to the pool fee, we differentiate \( \mathcal{L} \) with respect to \( f \):
\begin{equation}
    \frac{\partial{L}\left(f\right)}{\partial f}=-\frac{(r+1) \left(2 \left(-2 \Delta  \sqrt{f+1}+r+2\right)+3 f (r+2)\right)}{4 \Delta  \sqrt{f+1}}.
\end{equation}
Starting with \( f = 0 \) and given that \( \Delta > 1 + r \) (by Assumption \ref{ass:Delta}), the derivative at \( f = 0 \) is positive:
\begin{equation}
   \frac{\partial \mathcal{L}(f)}{\partial f}\bigg|_{f=0} = \frac{(r+1) (2 \Delta - r - 2)}{2 \Delta} > 0, 
\end{equation}
indicating that liquidity revenue increases with pool fee at this point. The derivative has roots:
\begin{equation}
f_{1,2} = \frac{-6 (r+2)^2 + 8 \Delta^2 \pm 4 \sqrt{4 \Delta^4 + 3 \Delta^2 (r+2)^2}}{9 (r+2)^2},    
\end{equation}
where the smallest root \( f_1 \) is negative and therefore not relevant. We need to show that the largest root \( f_2 \) is always positive, defining the threshold \( \overline{f} \).

For this, consider the numerator of $f_2$, labeled \( g(r, \Delta) \):
\[
g(r, \Delta) = 8 \Delta^2 + 4 \sqrt{4 \Delta^4 + 3 \Delta^2 (r+2)^2} - 6 (r+2)^2.
\]
This function has three roots in \( r \), all of which are negative: \( r = -2 \), \( r = -2(1 + \Delta) \), and \( r = 2(\Delta - 1) \). Since these roots are negative, for \( r \geq 0 \), \( g \) does not change sign and it is sufficient to examine \( g(0, \Delta) \):
\begin{equation}
    g(0, \Delta) = 8 \Delta^2 + 4 \sqrt{\Delta^2 (\Delta^2 + 3)} - 6.
\end{equation}
This is positive for any \( \Delta \geq 1 \), confirming that the largest root \( f_2 \) is positive and hence, \( \overline{f} \) exists and is positive. This completes the proof that the liquidity revenue increases with the pool fee until \( \overline{f} \) and decreases with pool fee for $f>\overline{f}$.
\end{proof}

\noindent \textbf{\large Lemma \ref{lem:advsel_fees}}
\begin{proof}
The cost of adverse selection for pool $k$ after evaluating equation \eqref{eq:AScost} is 
\begin{equation}
      \mathcal{A}\left(f_k\right)= v\frac{\left(\Delta -\sqrt{1+f}(1+r)\right) \left(\Delta ^2+\Delta  \sqrt{f+1}\left(1+r\right)+(f+1) (r-2) (r+1)\right)+(f+1)^{3/2} r^2 (r+1)}{3 \Delta }.
\end{equation}
We aim to demonstrate that \( \mathcal{A}(f) \) decreases as \( f \) increases. To do this, we calculate the partial derivative of \( \mathcal{A}(f) \) with respect to \( f \):
\begin{equation}
    \frac{\partial \mathcal{A}\left(f\right)}{\partial f}=\frac{(r+1) \left(-2 \Delta  \sqrt{f+1}+f (r+2)+r+2\right)}{2 \Delta  \sqrt{f+1}}<0
\end{equation}
The derivative is negative if $\Delta > \frac{1}{2} \sqrt{1+f} (2+r)$. Given that $\frac{1}{2}(2+r) < 1+r$, it follows that \( \mathcal{A}(f) \) decreases for any \( \Delta > (1+r) \sqrt{1+f} \), consistent with our assumption on \( \Delta \). \end{proof}

\noindent \textbf{\large Proposition \ref{prop:equilibria}}
\begin{proof}

First, consider the case in which \(\eta > \frac{\mathcal{L}\left(l\right)-\mathcal{L}\left(h\right)}{\mathcal{L}\left(l\right)-\mathcal{L}\left(h\right)+\mathcal{A}\left(l\right)-\mathcal{A}\left(h\right)}\) holds. This implies \(\qmg < 0\), and consequently, \(\pi_L - \pi_H < 0\) for all \(q\). Under this scenario, liquidity providers universally favor pool \(H\) over pool \(L\). They supply liquidity on pool \(H\) if and only if their participation constraint is satisfied, that is if \(q_i > \underline{q}_h\).

Conversely, if \(\eta \leq \frac{\mathcal{L}\left(l\right)-\mathcal{L}\left(h\right)}{\mathcal{L}\left(l\right)-\mathcal{L}\left(h\right)+\mathcal{A}\left(l\right)-\mathcal{A}\left(h\right)}\), then \(\qmg \geq 0\), allowing for a fragmented equilibrium. If \(\qmg \geq 0\), then \(\pi_L\) has a steeper slope compared to \(\pi_H\): profit increases more rapidly with liquidity supply in pool \(L\) than in pool $H$. There are two potential outcomes.

\begin{enumerate}
    \item \textbf{Dominance of pool $L$.} If \(0<\qmg < \underline{q}_\ell < \underline{q}_h\), as shown in the left-hand side panel of the diagram below, then the low-fee pool \(L\) captures the entire market share for any \(q_i\) that yields positive profits. The condition $\underline{q}_\ell < \underline{q}_h$ is equivalent to  
    \begin{equation}
        \underline{q}_\ell < \underline{q}_h \Leftrightarrow \frac{\mathcal{C}\left(h\right)}{\mathcal{C}\left(\ell\right)}>\frac{\left(1-\eta\right)\mathcal{L}\left(h\right)-\eta\mathcal{A}\left(h\right)}{\left(1-\eta\right)\mathcal{L}\left(\ell\right)-\eta\mathcal{A}\left(\ell\right)},
    \end{equation}
    which translates to $\eta>\frac{\mathcal{C}\left(\ell\right) \mathcal{L}\left(h\right)-\mathcal{C}\left(h\right) \mathcal{L}\left(\ell\right)}{\mathcal{C}\left(\ell\right) \left[\mathcal{L}\left(h\right)+\mathcal{A}\left(h\right)\right]-\mathcal{C}\left(h\right) \left[\mathcal{L}\left(\ell\right)+\mathcal{A}\left(\ell\right)\right]}$. Since we require $\eta\leq\frac{\mathcal{L}\left(\ell\right)}{\mathcal{L}\left(\ell\right)+\mathcal{A}\left(\ell\right)}$ by Assumption \ref{ass:eta}, it must be that $\mathcal{L}\left(\ell\right)\mathcal{A}\left(h\right)<\mathcal{L}\left(h\right)\mathcal{A}\left(\ell\right)$. 
    
    However, the parameter regions never overlap, ruling out this scenarios: we will show that $\mathcal{L}\left(\ell\right)\mathcal{A}\left(h\right)-\mathcal{L}\left(h\right)\mathcal{A}\left(\ell\right)>0$. To see this, we first note that $\eta \leq \frac{\mathcal{L}\left(l\right)-\mathcal{L}\left(h\right)}{\mathcal{L}\left(l\right)-\mathcal{L}\left(h\right)+\mathcal{A}\left(l\right)-\mathcal{A}\left(h\right)}$ is equivalent to:
    \begin{equation}\label{eq:c9}
        \eta\leq \frac{3 \left(h \left(2 \Delta -\sqrt{h+1} r-2 \sqrt{h+1}\right)-\ell \left(2 \Delta -\sqrt{\ell+1} r-2 \sqrt{\ell+1}\right)\right)}{(r+2) \left(\sqrt{h+1}\left(h-2 \right)-\sqrt{\ell+1}\left(\ell-2\right)\right)},
    \end{equation}
    which implies that $h \left(2 \Delta -\sqrt{h+1} r-2 \sqrt{h+1}\right)-\ell \left(2 \Delta -\sqrt{\ell+1} r-2 \sqrt{\ell+1}\right)>0$ since the denominator is always positive. We use the inequality in \eqref{eq:c9} and obtain that
    \begin{equation}
    \mathcal{L}\left(\ell\right)\mathcal{A}\left(h\right)-\mathcal{L}\left(h\right)\mathcal{A}\left(\ell\right)>\frac{1+r}{6\Delta^2} g,
    \end{equation}
    where 
    \begin{align}
        g&\equiv \underbrace{(1+r) l  \left(2 \Delta-\sqrt{l+1}\left(r+2\right)\right)}_{>0} \times \\
        & \times \left(\left(h-\ell\right)\Delta - (r+2) \left(\sqrt{h+1}-\sqrt{l+1}\right)+\underbrace{h  \left(2 \Delta-\sqrt{h+1}\left(r+2\right)\right)-l  \left(2 \Delta-\sqrt{l+1}\left(r+2\right)\right)}_{>0}\right)>0,
    \end{align}
    given Assumption \ref{ass:Delta} and equation \eqref{eq:c9}. To see that $\left(h-\ell\right)\Delta - (r+2) \left(\sqrt{h+1}-\sqrt{l+1}\right)>0$, we first note the expression increases in $\Delta$ and is therefore larger than $\sqrt{h+1} (r+1) (h-l)+(r+2) \left(\sqrt{l+1}-\sqrt{h+1}\right)$ for $\Delta>\sqrt{1+h}\left(1+r\right)$. The latter expression increases in $h$ and equals zero for $h=\ell$.The latter expression increases in $h$ and equals zero for $h=\ell$.  
    
    Therefore, there are no parameter values for which \(0<\qmg < \underline{q}_\ell < \underline{q}_h\) and \(\eta \leq \frac{\mathcal{L}\left(l\right)-\mathcal{L}\left(h\right)}{\mathcal{L}\left(l\right)-\mathcal{L}\left(h\right)+\mathcal{A}\left(l\right)-\mathcal{A}\left(h\right)}\), which rules out the case of pool $L$ attracting full market share.
    
    \item \textbf{Fragmented Market Equilibrium.} The right-hand side panel depicts the scenario \(\underline{q}_h < \underline{q}_\ell < \qmg\). Here, liquidity providers with \(q_i\) in the range \((\underline{q}_h, \qmg]\) achieve higher positive profits in pool \(H\), while those with \(q_i > \qmg\) obtain larger profits in pool \(L\). The condition $\underline{q}_h < \underline{q}_\ell$ is equivalent to  
    \begin{equation}
        \underline{q}_h \leq \underline{q}_\ell \Leftrightarrow \frac{\mathcal{C}\left(h\right)}{\mathcal{C}\left(\ell\right)}\leq\frac{\left(1-\eta\right)\mathcal{L}\left(h\right)-\eta\mathcal{A}\left(h\right)}{\left(1-\eta\right)\mathcal{L}\left(\ell\right)-\eta\mathcal{A}\left(\ell\right)},
    \end{equation}
    which is always true if $\eta \leq \frac{\mathcal{L}\left(l\right)-\mathcal{L}\left(h\right)}{\mathcal{L}\left(l\right)-\mathcal{L}\left(h\right)+\mathcal{A}\left(l\right)-\mathcal{A}\left(h\right)}$ as seen above.
\end{enumerate}

\begin{center}
    \begin{tikzpicture}[scale=1.5]

    \draw[thick,->] (0,-1) -- (0,2.2) node[anchor=south west] {$\pi_k$};
    \draw[thick,->] (0,0) -- (3,0) node[anchor=north] {q};

    \draw[blue, thick](0,-0.5) -- (3,2);
    \node[blue, align=center] at (0.55,0.2) {$\underline{q}_\ell$};
    \node[blue, align=center] at (3.2,2.1) {$\pi_L$};
    \node[blue, align=center] at (-0.2,-0.5) {$\mathcal{C}_\ell$};

    \draw[red, thick](0,-0.3) -- (3,0.5);
    \node[red, align=center] at (1.3,0.2) {$\underline{q}_h$};
    \node[red, align=center] at (3.2,0.6) {$\pi_H$};
    \node[red, align=center] at (-0.2,-0.3) {$\mathcal{C}_h$};

    \node[black, align=center] at (0.47,-0.3){$\qmg$};

    \node[align=center] at (1.5, 3) {(a) Single pool market (with fee $\ell$)};

    \draw[thick,->] (5,-1) -- (5,2.2) node[anchor=south west] {$\pi_k$};
    \draw[thick,->] (5,0) -- (8,0) node[anchor=north] {q};

    \draw[blue, thick](5,-0.8) -- (8,2);
    \node[blue, align=center] at (6,-0.2) {$\underline{q}_\ell$};
    \node[blue, align=center] at (8.2,2.1) {$\pi_L$};
    \node[blue, align=center] at (4.8,-0.8) {$\mathcal{C}_\ell$};

    \draw[red, thick](5,-0.2) -- (8,1);
    \node[red, align=center] at (5.4,0.2) {$\underline{q}_h$};
    \node[red, align=center] at (8.2,1.1) {$\pi_H$};
    \node[red, align=center] at (4.8,-0.2) {$\mathcal{C}_h$};

    \node[black, align=center] at (6,0.4) {$\qmg$};

        \node[align=center] at (6.5, 3) {(b) Fragmented market};

\end{tikzpicture}
\end{center}

It is crucial to note that configurations where \(\underline{q}_\ell < \qmg < \underline{q}_h\) or \(\underline{q}_h < \qmg < \underline{q}_\ell\) are not feasible, as they would lead to a contradiction where profits are simultaneously positive in one pool and negative in the other at the indifference point \(\qmg\).

\end{proof}

\noindent \textbf{\large Proposition \ref{cor:comp_stat_ms}}
\begin{proof}
We first note that both $\qmg$ and $\underline{q}_h$ scale linearly with $\Gamma$: that is, there exists $Q_t>Q_h>0$ such that $\qmg=\Gamma Q_t$ and $\underline{q}_h=\Gamma Q_h$ where $Q_t$ and $Q_h$ are not functions of $\Gamma$. Next, we compute the partial derivative of $w_\ell$ with respect to $\Gamma$
\begin{equation}
\frac{\partial w_\ell}{\partial \Gamma}=\frac{\Gamma  (Q_h-Q_t) e^{\frac{\Gamma  (Q_h-Q_t)}{\lambda }} (\Gamma  Q_h Q_t+\lambda  (Q_h+Q_t))}{\lambda  (\lambda +\Gamma  Q_h)^2}<0,
\end{equation}
since $Q_h<Q_t$ and all other terms are positive.\end{proof}

\noindent \textbf{\large Proposition \ref{prop:optimality}}
\begin{proof}
The two-pool gains from trade for an $\LT$ with private value $v\left(1+\delta\right)$ are
\begin{align}
    \text{GainsFromTrade}\left(\left\{h,\ell \mid \delta \right\}\right)&=v \delta T_H\min\left\{1,\frac{1+r}{r}\max\left\{0,1-\sqrt{\frac{1+h}{1+\delta}}\right\}\right\}+\nonumber \\&+v \delta T_L\min\left\{1,\frac{1+r}{r}\max\left\{0,1-\sqrt{\frac{1+\ell}{1+\delta}}\right\}\right\}.
\end{align}
We set $h=f$ such that the marginal $\LP$ entering the market is the same as in the single-fee pool; that is, $T_H+T_L=e^{-q_h\lambda}\left(q_h+\lambda\right)$. Since $\min\left\{1,\frac{1+r}{r}\max\left\{0,1-\sqrt{\frac{1+f}{1+\delta}}\right\}\right\}$ decreases in $f$, it follows that:
\begin{align}
    \text{GainsFromTrade}\left(\left\{h,\ell \mid \delta \right\}\right)&\geq v\delta \underbrace{\left(T_H+T_L\right)}_{=e^{-q_h\lambda}\left(q_h+\lambda\right)}\min\left\{1,\frac{1+r}{r}\max\left\{0,1-\sqrt{\frac{1+h}{1+\delta}}\right\}\right\}\\
    &=\text{GainsFromTrade}\left(\left\{h\right\}\mid \delta \right\},
\end{align}
with strict inequality if $\qmg<Q$ such that the low-fee pool attracts a positive mass of $\LP$s. The inequality holds for any $\delta$, and therefore it remains true if we aggregate the gains from trade over the distribution $\LT$ private values. We note that the expected gains from trade per unit of liquidity is
\begin{align}
    &\mathbb{E}\min\left\{1,\frac{1+r}{r}\max\left\{0,1-\sqrt{\frac{1+h}{1+\delta}}\right\}\right\}- \nonumber \\&=\frac{6 \sqrt{f+1} (r+1) \log (r+1)+r \left(2 \left(\Delta ^3-3 \Delta -\sqrt{f+1}\right)+\sqrt{f+1} (f (r+1) (r+2)+r (r+3))\right)}{6 \Delta  r}>0,
\end{align}
which decreases in $f$ since
\begin{align}
    \frac{\partial \mathbb{E}\min\left\{1,\frac{1+r}{r}\max\left\{0,1-\sqrt{\frac{1+h}{1+\delta}}\right\}\right\}}{\partial f}=-\frac{(r+1) ((f+1) r (r+2)-2 \log (r+1))}{4 \Delta  \sqrt{f+1} r}<0.
\end{align}

\end{proof}

\newpage
\section{Just-in-time liquidity \label{sec:app-jit}}
Just-in-time (JIT) liquidity is a strategy that leverages the transparency of orders on the public blockchains. If a liquidity provider observes an incoming large order that has not been processed by miners and it deems uninformed in the public mempool, it can conveniently re-arrange transactions and propose a sequence of actions to sandwich this trade as follows:
\begin{enumerate}
    \item Add a large liquidity deposit at block position $k$, at the smallest tick around the current pool price. 
    \item Let the trade at block position $k+1$ execute and receive liquidity fees.
    \item Remove or burn any residual un-executed liquidity at block position $k+2$.
\end{enumerate}
The mint size is optimally very large (i.e., of the order of hundred of millions USD for liquid pairs), such that the JIT liquidity provider effectively crowds out the existing liquidity supply and collects most fees for the trade. That is, the strategy is made possible by pro-rata matching on decentralized exchanges because with time priority, the JIT provider cannot queue-jump existing liquidity providers. Since the JIT liquidity provider does not want to passively provide capital, it removes any residual deposit immediately after the trade.

We identify JIT liquidity events by the following algorithm as in \citet{WanAdams2022}:
\begin{enumerate}
\item Search for mints and burns in the same block, liquidity pool, and initiated by the same wallet address. The mint needs to occur exactly two block positions ahead of the burn (at positions $k$ and $k+2$).
\item Classify the mint and the burn as a JIT event if the transaction in between (at position $k+1$) is a trade in the same liquidity pool.
\end{enumerate}
 
JIT events are rare in our sample, and account for less than 1\% of the traded volume on Uniswap v3. Further, more than half of them occur in a single pair - USDC-WETH, and in low-fee pools. The Uniswap Labs provides further discussions on the aggregate impact of JIT liquidity provision \href{https://uniswap.org/blog/jit-liquidity}{here}. Regarding the economic effects, JIT liquidity reduces price impact for incoming trades, but dilutes existing liquidity providers in the pro-rata markets, and can discourage liquidity supply in the long run.

\newpage
\section{Impermanent loss measure \label{app:IL}}
We build our measure of impermanent loss in line with the definition of token reserves within a price range in the Uniswap V3 white paper \citep{Uniswapv3Core2021} and Section 4.1 in \citet{Heimbach2023}.

Consider a liquidity provider who supplies $L$ units of liquidity into a pool trading a token $x$ for a token $y$. The chosen price range is $\left[p_\ell, p_u\right]$ with $p_\ell<p_u$. Further, the current price of the pool is $p_0$. We are interested in computing the impermanent loss at a future point in time, when the price updates to $p_1$.

From \citet{Uniswapv3Core2021}, the actual amount of tokens $x$ and $y$ (``real reserves'') deposited on a Uniswap v3 liquidity pool with a price range $\left[p_\ell, p_u\right]$ to yield liquidity $L$ are functions of the current pool price $p$:
\begin{equation}\label{eq:reserves_t0}
    x\left(p\right)=\begin{cases}
        L \times \left(\frac{1}{\sqrt{p_\ell}}-\frac{1}{\sqrt{p_u}}\right) & \text { if } p\leq p_\ell \\
        L \times \left(\frac{1}{\sqrt{p}}-\frac{1}{\sqrt{p_u}}\right) & \text { if } p_\ell<p\leq p_u \\
        0 & \text { if } p > p_u
    \end{cases} \text{ and }     y\left(p\right)=\begin{cases}
        0 & \text { if } p\leq p_\ell \\
        L \times  \left(\sqrt{p}-\sqrt{p_\ell}\right) & \text { if } p_\ell<p\leq p_u \\
        L \times  \left(\sqrt{p_u}-\sqrt{p_\ell}\right) & \text { if } p > p_u.
    \end{cases}
\end{equation}

From equation \eqref{eq:reserves_t0}, the value of the liquidity position at $t=1$ is therefore
\begin{equation}
    V_\text{position}=p_1 x\left(p_1\right) + y\left(p_1\right)=\begin{cases}
        L p_1 \times \left(\frac{1}{\sqrt{p_\ell}}-\frac{1}{\sqrt{p_u}}\right) & \text { if } p_1\leq p_\ell \\
        L \times \left(2\sqrt{p_1}-\frac{p_1}{\sqrt{p_u}}-\sqrt{p_\ell}\right) & \text { if } p_\ell<p_1\leq p_u \\
         L \times  \left(\sqrt{p_u}-\sqrt{p_\ell}\right) & \text { if } p_1 > p_u.
    \end{cases} 
\end{equation}

Conversely, the value of a strategy where the liquidity provider holds the original token quantities and marks them to market at the updated price is
\begin{equation}
    V_\text{hold}=p_1 x\left(p_0\right) + y\left(p_0\right)=\begin{cases}
        L p_1 \times \left(\frac{1}{\sqrt{p_\ell}}-\frac{1}{\sqrt{p_u}}\right) & \text { if } p_0\leq p_\ell \\
        L \times \left(\frac{p_1+p_0}{\sqrt{p_0}}-\frac{p_1}{\sqrt{p_u}}-\sqrt{p_\ell}\right) & \text { if } p_\ell<p_0\leq p_u \\
         L \times  \left(\sqrt{p_u}-\sqrt{p_\ell}\right) & \text { if } p_0 > p_u.
    \end{cases} 
\end{equation}

The impermanent loss is then defined as the excess return from holding the assets versus providing liquidity on the decentralized exchange:
\begin{equation}
    \text{ImpermanentLoss}=\frac{V_\text{hold}-V_\text{position}}{V_\text{hold}}.
\end{equation}

Empirically, we follow \citet{Heimbach2023} and compute impermanent loss for ``symmetric'' positions around the current pool price, that is $p_\ell=p_0 \alpha^{-1}$ and $p_u= p_0\alpha$, with $\alpha>1$. We allow for a time lag of one hour between $p_0$ and $p_1$.


\newpage

\newpage
\end{document}